\documentclass[preprint,aps,superscriptaddress,preprintnumbers,amsmath,amssymb]{revtex4-1}
\usepackage{graphicx}
\usepackage[version=2]{mhchem}
\usepackage{subcaption}
\usepackage{array}
\usepackage{siunitx}
\usepackage{booktabs}
\usepackage{gensymb}
\usepackage{multirow}
\usepackage{dcolumn} 
\usepackage{setspace}
\usepackage{array}
\usepackage{psfrag}
\usepackage{textcomp}
\usepackage{color}
\usepackage{makecell}
\usepackage[thinlines]{easytable} 

\newcommand{\eqn}[1]{\mbox{Eq.\hspace{1pt}(\ref{#1})}}

\newcommand{\eqtn}[2]{\begin{equation} \label{#1} #2 \end{equation}}

\def\br{{\mathbf{r}}}

\def\dscf{{$\Delta$SCF}}
\def\txcdft{{$\tau$XCDFT}}

\graphicspath{ {figures/} }

\begin{document}

\title{Capturing Multireference Excited States by Constrained DFT}

\author{Nell Karpinski}

\affiliation{Department of Chemistry, Rutgers University, Newark, NJ 07102, USA}

\author{Pablo Ramos} 
\email{p.ramos@rutgers.edu}

\affiliation{Department of Chemistry, Rutgers University, Newark, NJ 07102, USA}

\author{Michele Pavanello}
\email{m.pavanello@rutgers.edu}

\affiliation{Department of Chemistry, Rutgers University, Newark, NJ 07102, USA}
\affiliation{Department of Physics, Rutgers University, Newark, NJ 07102, USA}

\begin{abstract}
The computation of excited electronic states with commonly employed (approximate) methods is challenging, typically yielding states of lower quality than the corresponding ground state for a higher computational cost.
In this work, we present a mean field method that extends the previously proposed eXcited Constrained DFT (XCDFT) from single Slater determinants to ensemble 1-RDMs for computing low-lying excited states. The method still retains an associated computational complexity comparable to a semilocal DFT calculation while at the same time is capable of approaching states with multireference character. We benchmark the quality of this method on well-established test sets, finding good descriptions of the electronic structure of multireference states and maintaining an overall accuracy for the predicted excitation energies comparable to semilocal TDDFT.
\end{abstract}

\date{\today}
\maketitle

\section{Introduction}
Models of molecules and materials typically require the knowledge of excited electronic states and must be able to approach complex dynamical regimes. For example, in energy sciences and photochemistry, often the dynamics involve interaction with external electromagnetic fields or require to characterize states that are very close in energy. Thus, the task at hand is formulating a computationally efficient model of electronic excited states capable of handling the many difficult cases that, unfortunately, routinely arise.

Density Functional Theory (DFT) has been the workhorse of electronic structure theory for the computation of excited electronic states and their dynamics {\it via} its time-dependent extension (TDDFT). Unfortunately, TDDFT has some notable shortcomings when it is implemented in the adiabatic and the semilocal density approximations. Conical intersections, charge transfer states and Rydberg states are among those cases where practical implementations of TDDFT struggle to provide a physical model. More recently, multiconfigurational DFT methods, such as ensemble DFT \cite{Filatov_2014,Filatov_2017,Yang_2017,Deur_2019,Franck_2013}, constrained DFT \cite{evan2013,ramo2018a,kadu2012}, block-localized DFT \cite{Cembran_2010,cem2009}, DFT/MRCI \cite{Grimme_1999}, and even flavors of ground state DFT \cite{Mei_2019} have been proposed as innovative protocols for extracting excitation energies in a computationally efficient way while still making use of density functionals in their formulation. 

Constrained DFT \cite{vanv2010a} is particularly interesting because it does not need an active space and, instead, targets directly the excited states with the wanted character \cite{wu2006}. Traditionally it has been employed for generating charge and spin-localized states (diabatic states). However, recent works including our own have borrowed the general constrained DFT idea and proposed methods for computing valence excited states \cite{evan2013,ramo2018a,zieg2016,ziegl2014,Ziegler2012}. 

In this work, we continue the development of the eXcited Constrained DFT (XCDFT) method\cite{ramo2018a}. In XCDFT, a variational procedure produces excited states energies and densities of similar quality to the ground states ones for a similarly comparable computational cost. In essence, XCDFT exploits the machinery of ground state Kohn-Sham DFT for the generation of excited states \cite{Ayers_2009,Ayers_2012,Levy_1999}. Inspired by density functional perturbation theory \cite{Baroni_1987}, XCDFT does not require the use of unoccupied bands (virtuals) as it resolves the space of virtuals by projection. The Fock operator is then augmented by a nonlocal and orbital dependent constraining potential exerting a force on the electrons, leading to a selfconsistent solution for the targeted excited state. XCDFT is similar in spirit to \dscf\ without the inconvenience of incurring in variational collapses. In our previous publication \cite{ramo2018a}, we carried out a careful comparison of XCDFT against \dscf, and linear-response semilocal TDDFT and found that its accuracy compares to them (about 0.5 to 1.0 eV deviation from benchmark values for the chosen test set). 

Unfortunately and similar to \dscf, due to the fact that XCDFT makes use of a single Slater determinant, when approaching degenerate excited states it fails to produce correct electronic structures. This is problematic because degenerate electronic excited states are ubiquitous.

In this work, we take inspiration from ensemble DFT methods and propose the use of thermal ensemble one-body reduced density matrices (1-RDMs) for describing the electronic structure of the excited state. We dub the resulting method \txcdft. This allows us to partially occupy excited state's Kohn-Sham orbitals and reach an accurate depiction of a multireference excited state at merely the expense of needing to compute a larger number of occupied orbitals. 

The paper is organized as follows: we first describe the theory and implementation of \txcdft\ and clearly show the involved approximations. After a short description of the computational details, needed for the reproducibility of the results, we show results on the same test set considered previously \cite{ramo2018a}, as well as additional complex large molecules, such as anthracene, tetracene and fullerene.
These additional large molecular systems are included because the description of their excited states' electronic structure may require more than single excitations from the reference determinant. Due to the variational nature of \txcdft, the orbitals are relaxed to infinite order making up most of the relaxation effects that are captured by multiple excitations in those methods exploiting a reference determinant. 

\section{Theory and Background}
The starting point of an XCDFT calculation is a reference ground state (gs) obtained from a regular KS calculation. From that, a projection operator $\hat{P}_o^g$ over the occupied space of gs, $\{|i_g\rangle\}$, is constructed. Namely,
\eqtn{dg}{\hat{P}_o^g=\sum_{i_g=1}^{\text{occ}}|i_g\rangle\langle i_g|.}

The electronic excitations are obtained by applying a nonlocal potential, $\hat{W}_c$, whose action is to ``fish out'' an electron into the virtual space of the reference gs. In the basis of the atomic orbitals (customarily indicated by Greek letters, $\nu$ and $\mu$), such potential is written as:
\eqtn{wcao}{\left( \hat{W}_c \right)_{\mu\nu}=\langle\mu | \hat 1 - \hat{P}_o^g | \nu\rangle,}
which then is used to define the constraint that only one electron should be excited to the virtual space of the reference gs,
\eqtn{vc}{1= \sum_{j=1}^\text{occ}\langle j_e | \hat{1} - \hat{P}_o^g | j_e \rangle \equiv \mathrm{Tr}\left[\hat{W}_c\hat{\gamma}_e\right]= N_e - \sum_{i_g,j_e=1}^\text{occ}\langle j_e | i_g \rangle\langle i_g | j_e \rangle.}
Where $\hat{\gamma}_e$ is the density matrix of the excited state, $N_e$ is the total number of electrons, and $V_c$ is an appropriate constant (a Lagrange multiplier that ensures the constraint is satisfied) and $ | j_e \rangle$ are the excited state occupied orbitals. In this context, $V_c$ is the value of the excitation energy and needs to be determined selfconsistently.

XCDFT yields excitation energies in semiquantitative agreement with TDDFT and benchmark calculations, however, we noticed \cite{ramo2018a} that whenever it is required to go beyond a single Slater determinant, spurious contributions from more than singly excited configuration state functions arise degrading the excited state's electronic structure. One particularly deteriorating factor is the resulting significant overlap with the gs KS wavefunction. As this problem only arises when multreference excited states are considered, we turned to the several studies carried out to understand and deal with static correlation in Kohn-Sham DFT\cite{Lieb_1983,Schipper_1998,Gritsenko_1997}. It is known that when near degeneracies arise (typical case of a multireference system), an ensemble of noninteracting electrons provides a more convenient reference than typical single Slater determinants\cite{Wang_1996,Schipper_1998}. Thus, in this work, we allow XCDFT excited states to probe finite-temperature ensemble 1-RDMs as follows:
\eqtn{denT}{\hat{\gamma}_e = \sum_{i_e} |i_e \rangle f_i \langle i_e |,}
with $f_i$ are the occupation numbers which are determined by the Fermi--Dirac distribution function,
\eqtn{fdd}{f_j \equiv f(\epsilon_j - \mu) = \left[1 + \text{exp}(\beta (\varepsilon_j - \mu))\right]^{-1},}
with $\beta = \frac{1}{k_B\tau}$ (a parameter of the method), $\mu$ is the chemical potential, and $\varepsilon_j$ are the orbital energies. Smearing the orbital occupations is a well-known strategy that has been used in mean-field calculations\cite{Slater_1969} of both finite and periodic systems when degeneracies appear.   

Thus, \eqn{vc} is modified to
\eqtn{vc_tau}{1= \sum_{j=1}^\text{occ}\langle j_e | \hat{1} - \hat{P}_o^g | j_e \rangle \equiv \mathrm{Tr}\left[\hat{W}_c\hat{\gamma}_e\right]= N_e - \sum_{i_g}^{occ}\sum_{j_e}^{\infty} \langle i_g | j_e \rangle f_{j_e}\langle j_e | i_g \rangle.}
We dub the resulting method \txcdft.

\section{Computational Details}
All XCDFT and \txcdft\ excited state calculations are performed with a development version of the Amsterdam Density Functional (ADF) 2019 program\cite{ADF2019authors}. To assess the performance of \txcdft, we consider the lowest excited state for a set of fifteen molecules \cite{ramo2018a} with the addition of the anthracene, tetracene and fullerene. As describe above (see theory section) we rely on the approximation that the smearing provided by the Fermi--Dirac distribution function is sufficient to account for the fractional occupations resulting from the multireference character of certain excited states. This smearing can be achieved by inducing an electronic temperature, which for sake of consistency was set to 500 K throughout all \txcdft\ calculations. The GGA functional PBE\cite{PBEc}, and the metaGGA functionals M06L\cite{zhao2008}, SCAN\cite{Sun_2015} and revTPSS\cite{PErdew_2009} are employed across the entire study along with the TZP basis set. We report XCDFT and \txcdft\ excitation energies by using the value of the corresponding $V_c$ Lagrange multiplier. In addition, the differential densities obtained with XCDFT are compared against the ones obtained from TDDFT, calculated with ORCA \cite{neese2012}. 

\section{Results and Discussion}
\subsection{Quality of the electron density}

We carried out an analysis of the electronic densities comparing the differential densities (i.e., the density difference between the excited state and ground state densities, $\Delta(\mathbf{r}) = \rho_e(\mathbf{r}) - \rho_g(\mathbf{r})$) obtained form \txcdft, XCDFT, TDDFT and EOM-CCSD. In Figure \ref{DDeom}, $\Delta(\mathbf{r})$ is displayed for acrolein, benzene and fullerene. 

\def\wideas{0.15\textwidth}
\begin{figure}
\caption{\label{DDeom} Comparison of the computed differential densities for all DFT methods used against EOM-CCSD. The standard deviation, $\sigma$, of each density difference against EOM-CCSD densities (calculated using Gaussian \cite{g16}) are shown.}
\begin{center}
{\fontsize{11}{11}\selectfont
\begin{tabular}{ccccc}
        TDDFT & $\Delta$SCF & XCDFT & \txcdft & EOM-CCSD \\
        \hline
        0.081 & 0.069 & 0.058 & 0.057 & \\
        \includegraphics[width=\wideas]{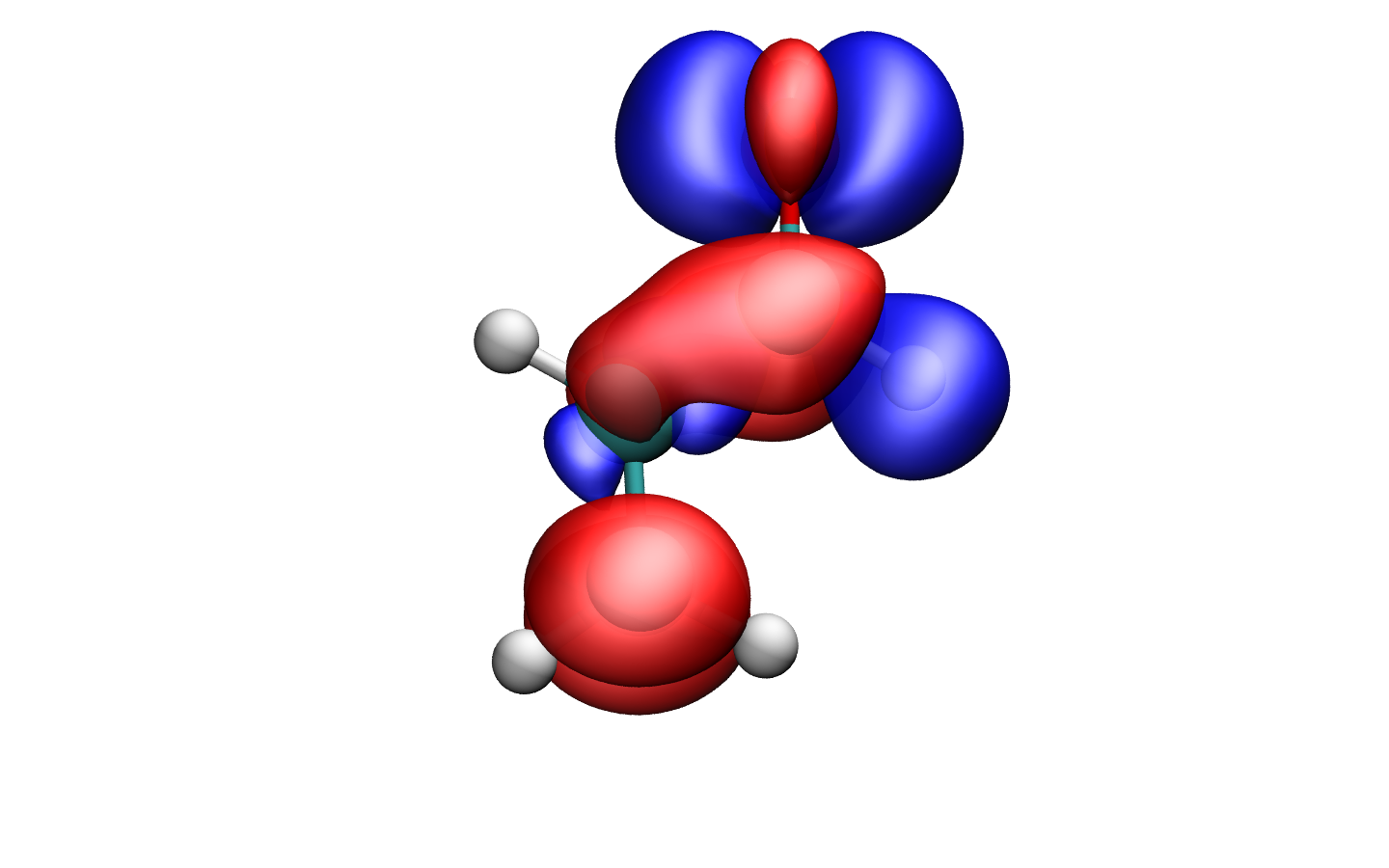}&\includegraphics[width=\wideas]{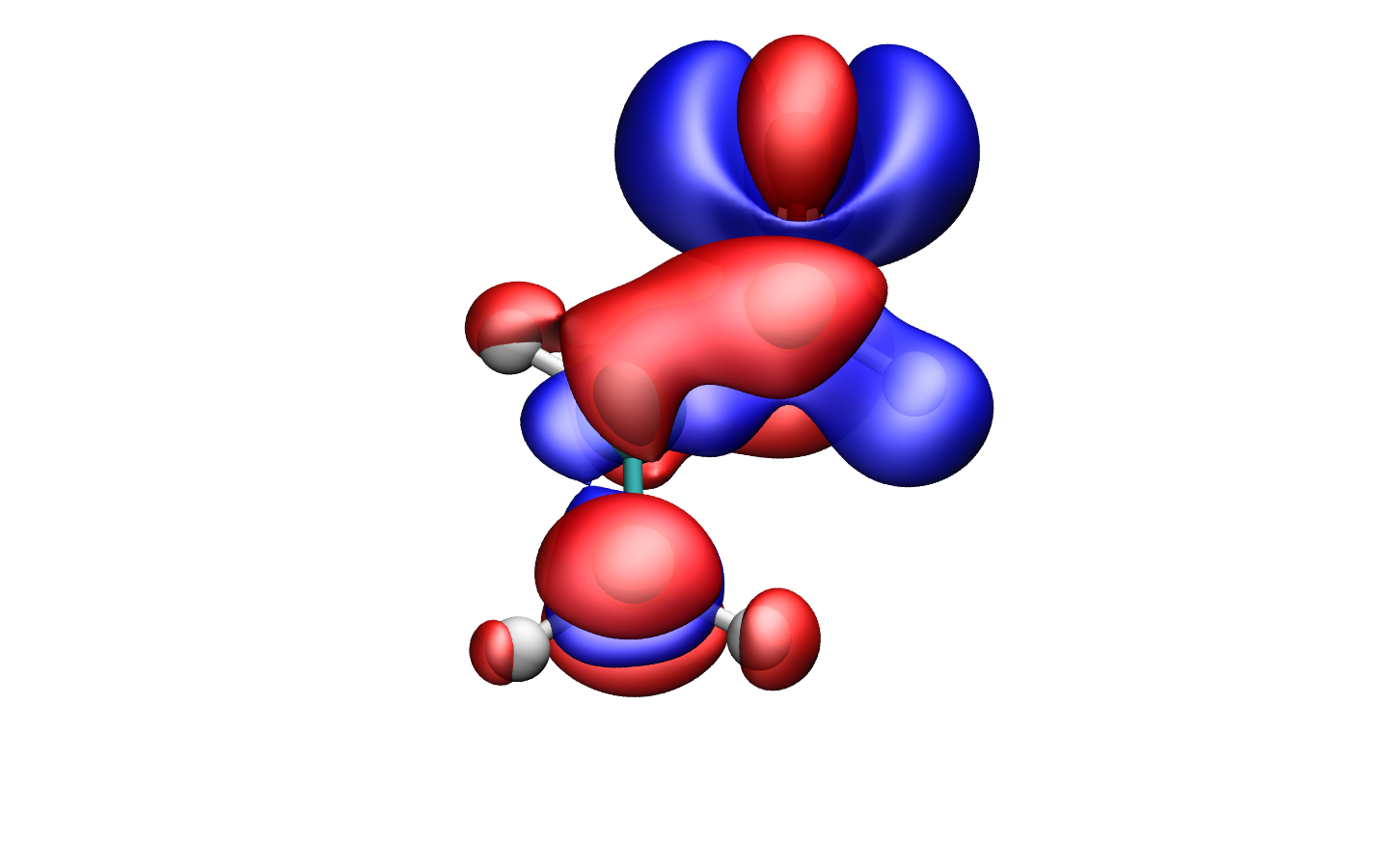}&\includegraphics[width=\wideas]{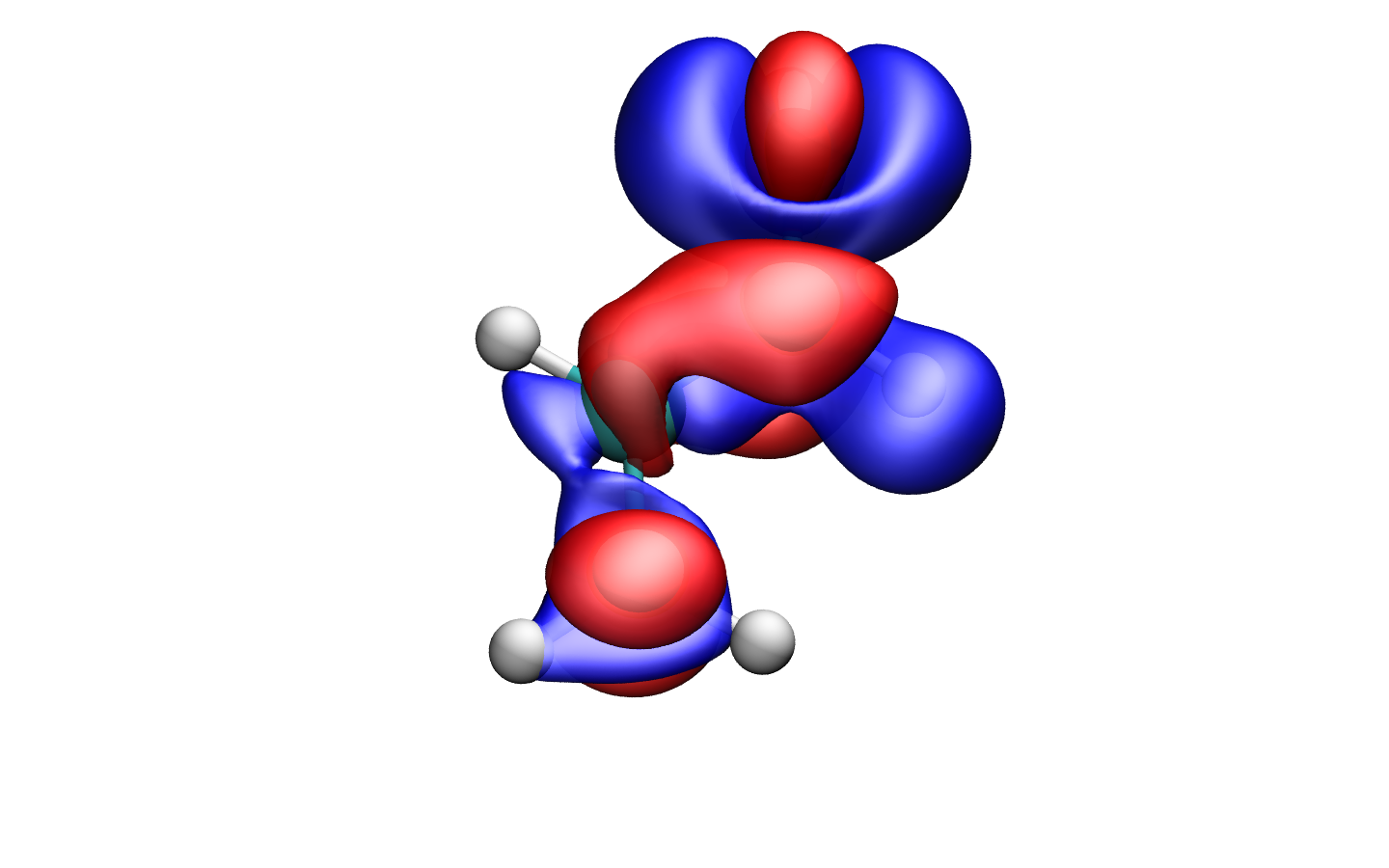}&\includegraphics[width=\wideas]{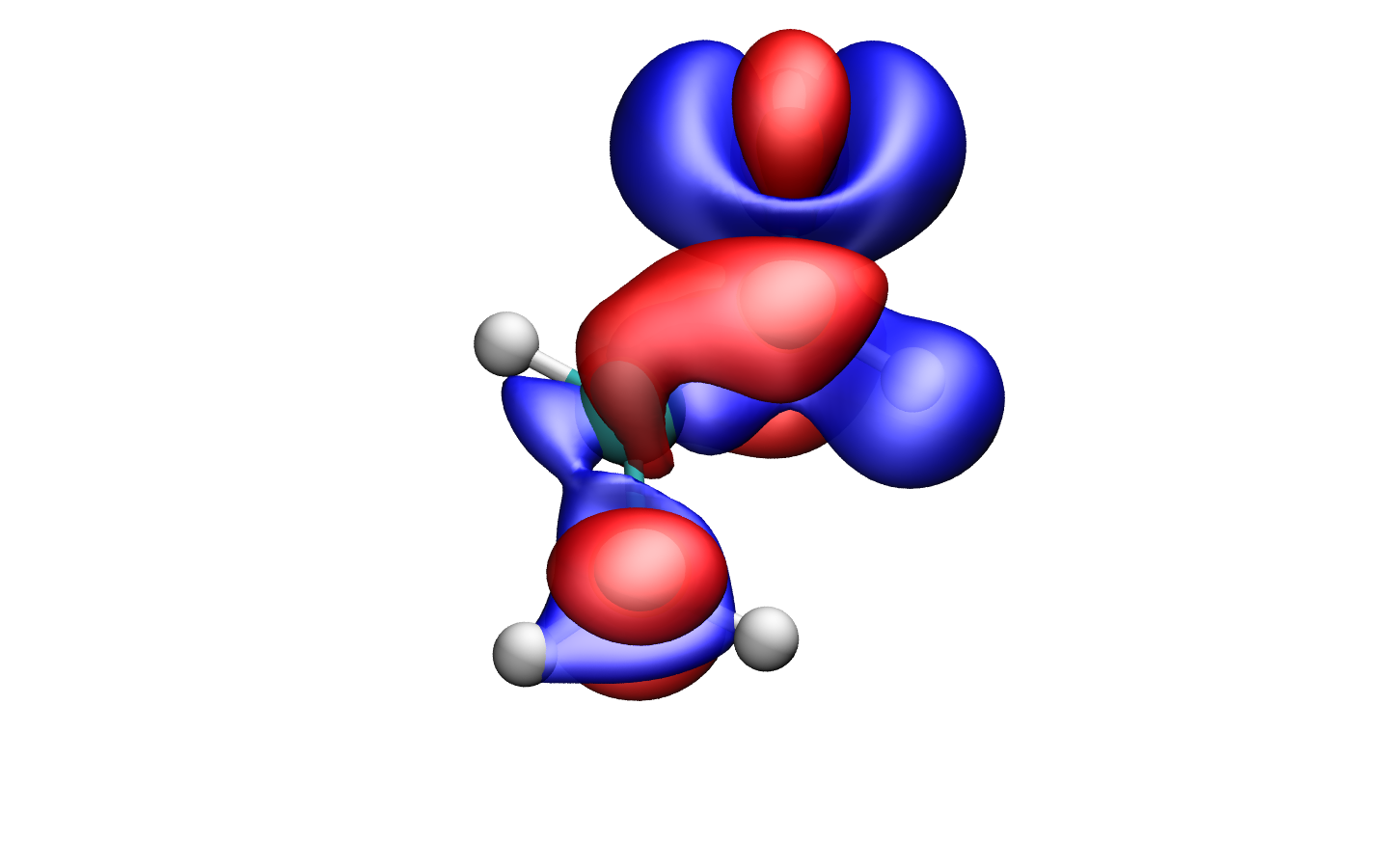}&\includegraphics[width=\wideas]{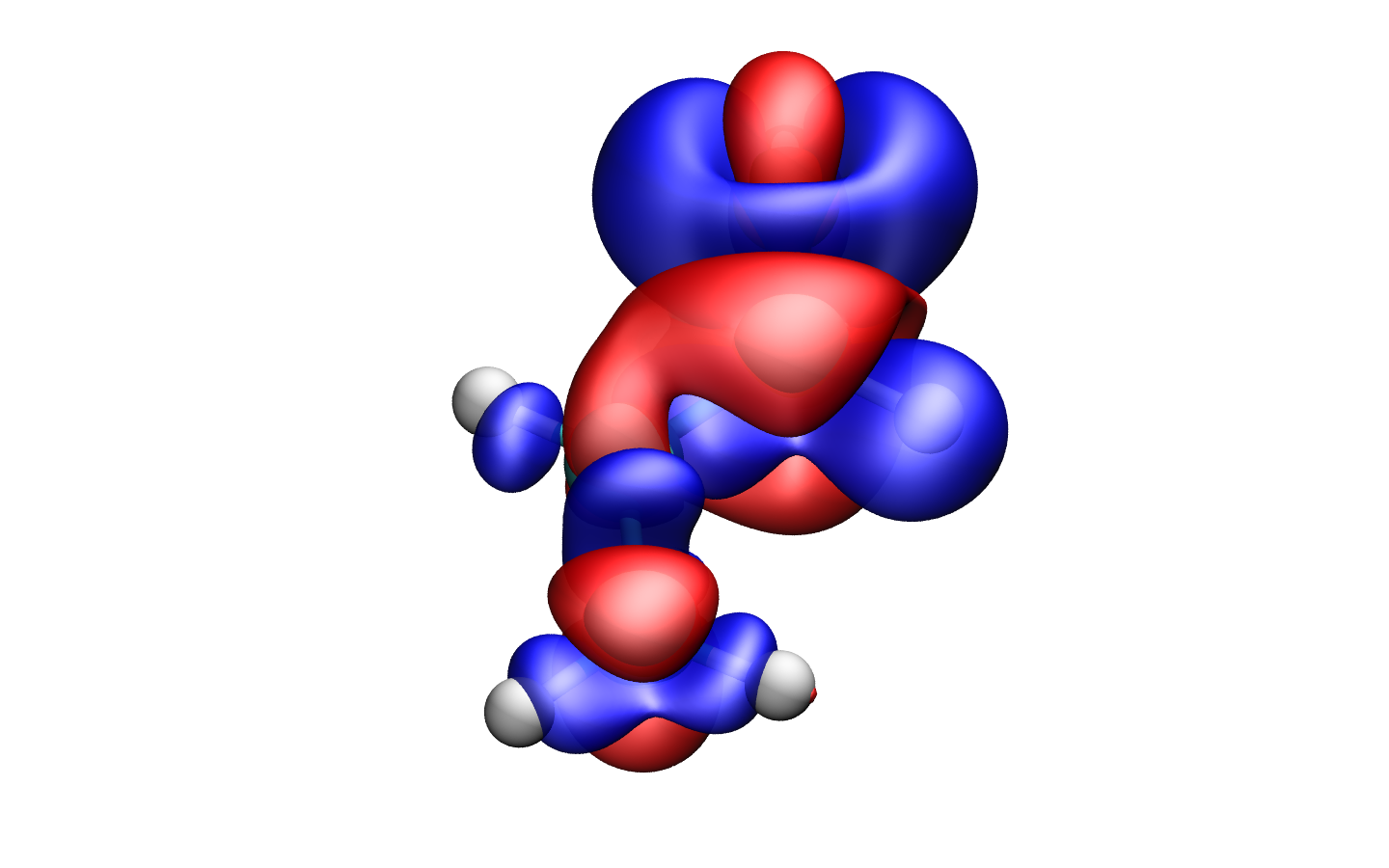}\\
        \hline
        0.053 & 0.066 & 0.063 & 0.033 & \\
        \includegraphics[width=\wideas]{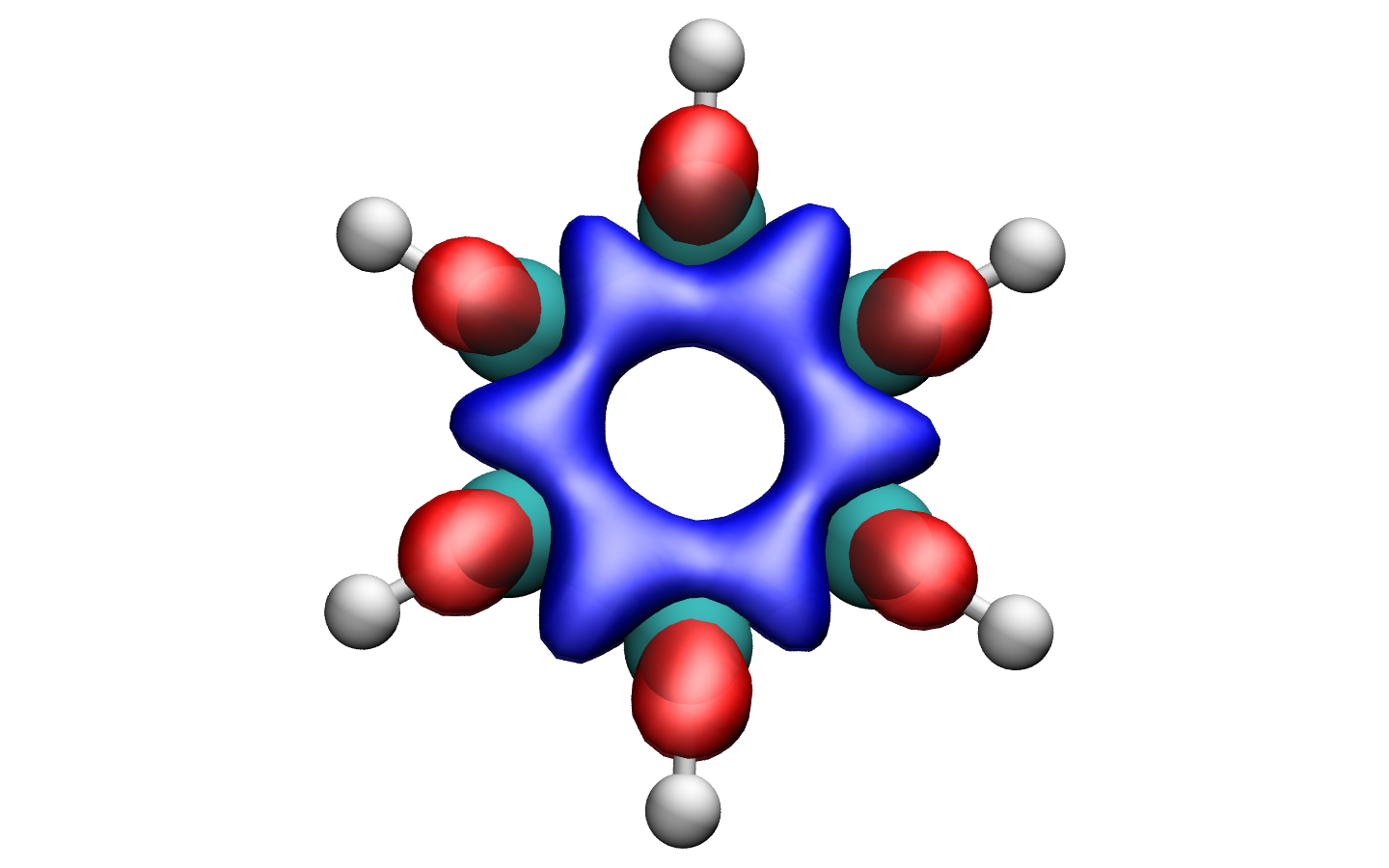}&\includegraphics[width=\wideas]{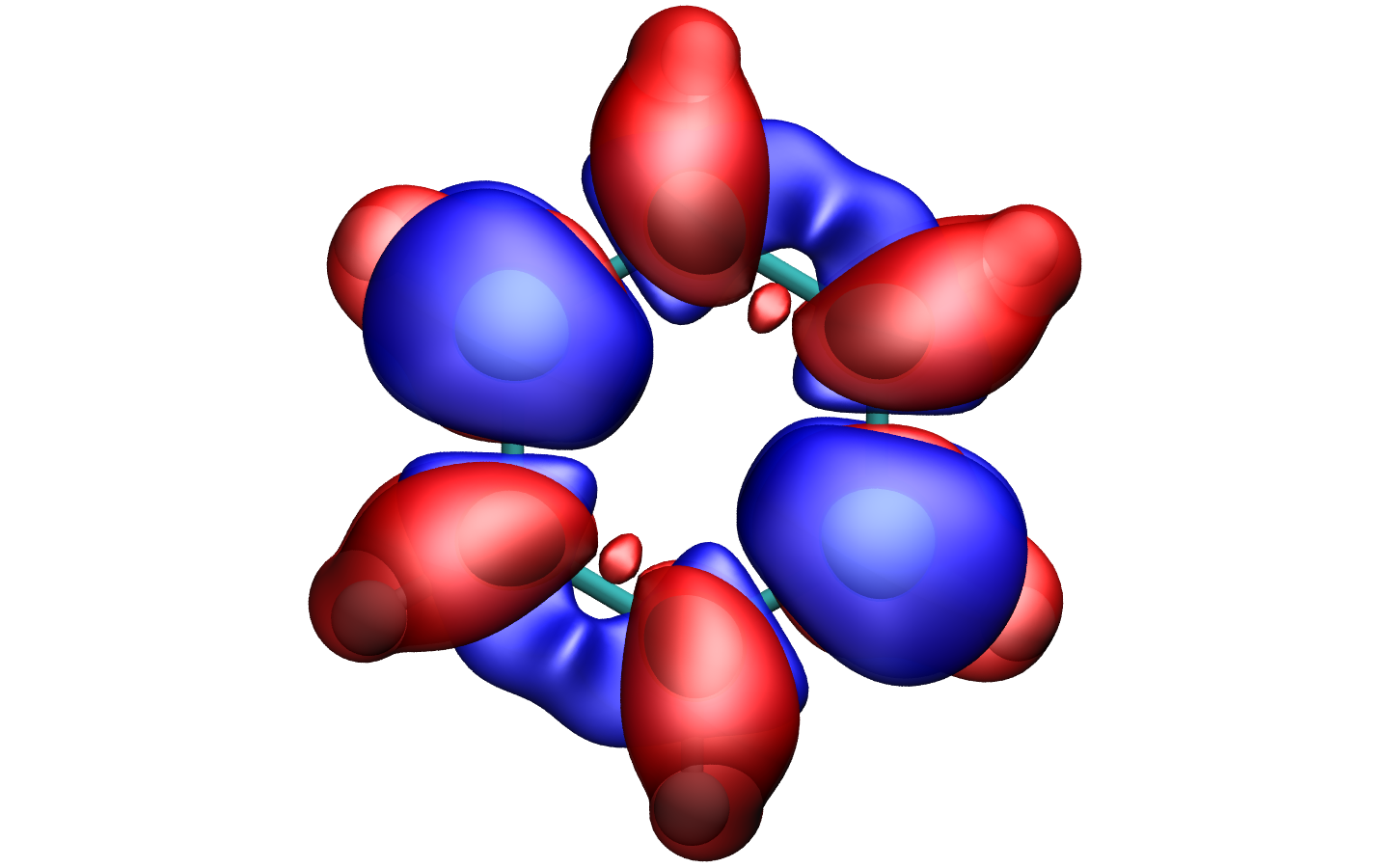}&\includegraphics[width=\wideas]{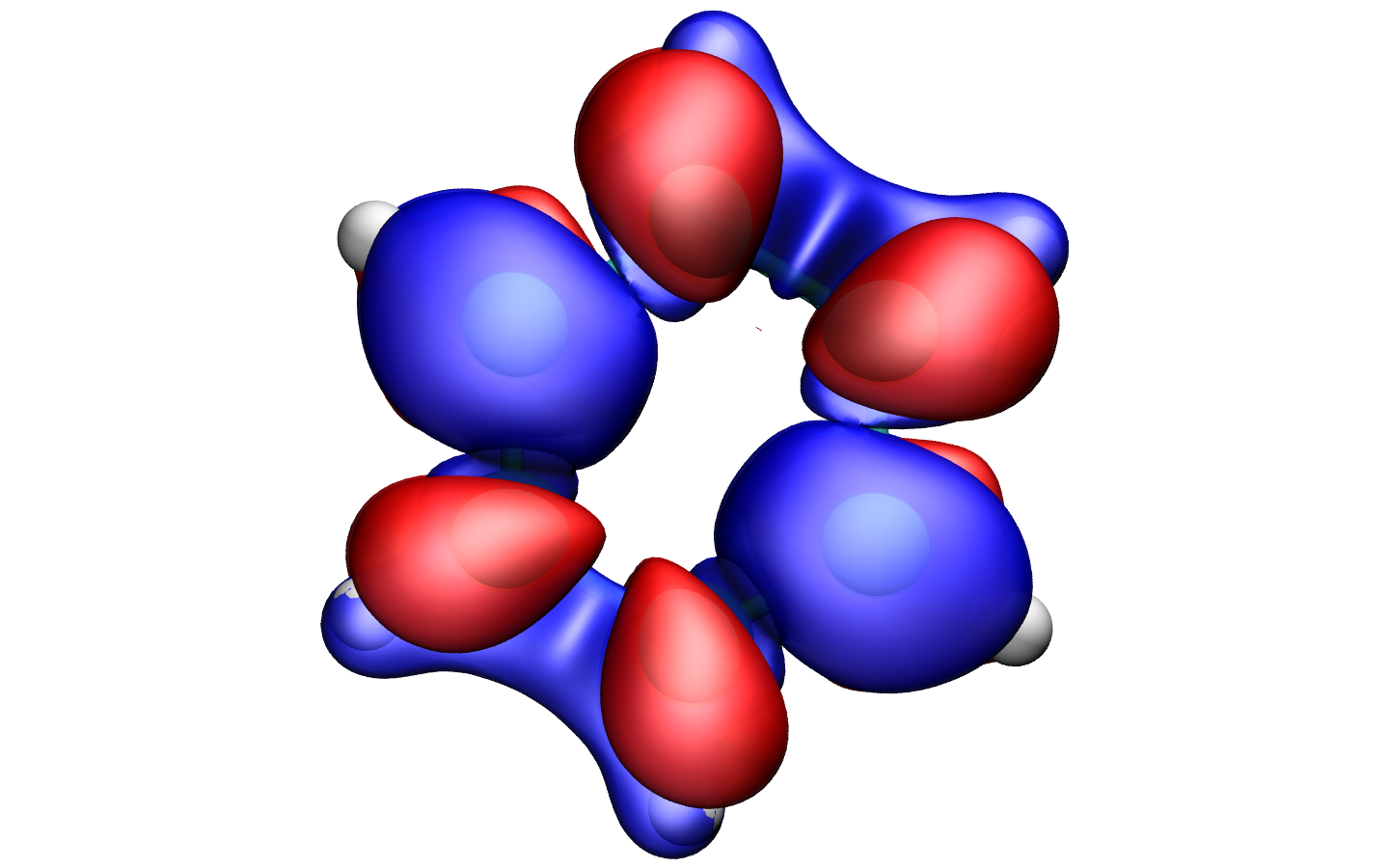}&\includegraphics[width=\wideas]{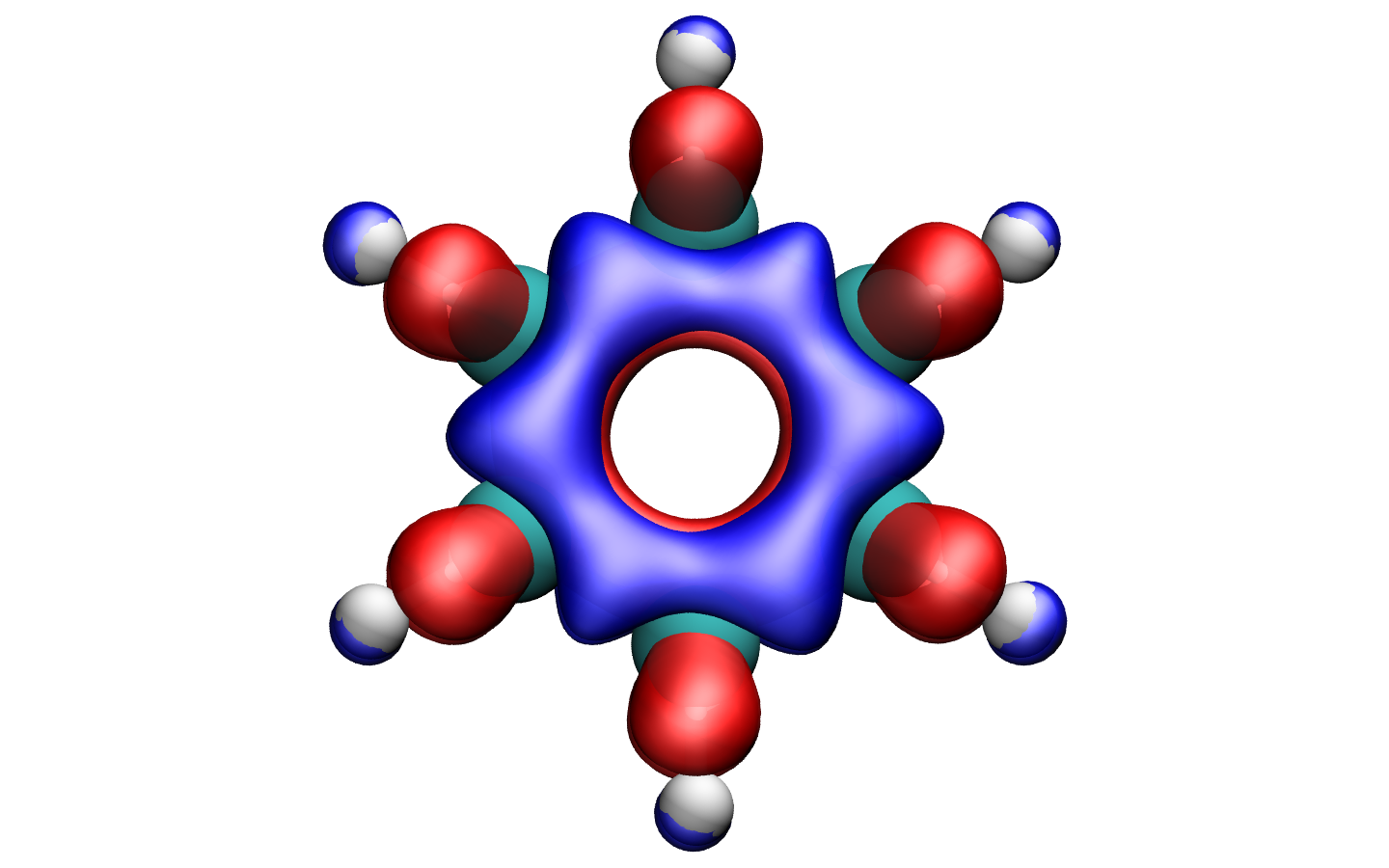}&\includegraphics[width=\wideas]{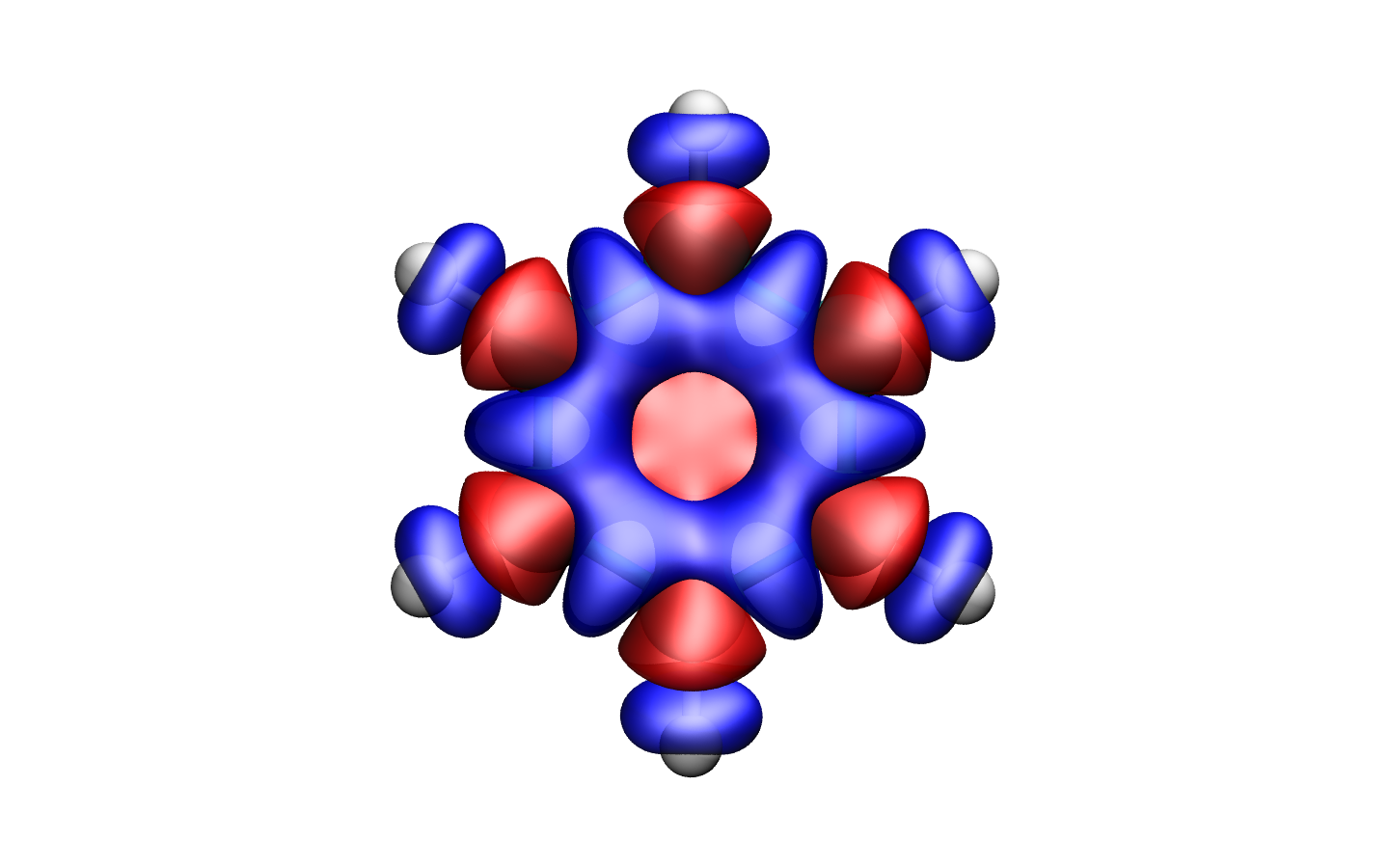}\\
\end{tabular}
}
\end{center}
\end{figure}

\def\wideas{0.25\textwidth}
\begin{figure}
\caption{\label{full} Fullerene density differences divided by contribution $\Delta(\br)>0$ in blue (top), and $\Delta(\br)<0$ in red (bottom). All isosurfaces are plotted with the same cutoff.}
\begin{center}
{\fontsize{11}{11}\selectfont
\begin{tabular}{ccc}
	TDDFT & XCDFT & \txcdft \\
	\hline
	\includegraphics[width=\wideas]{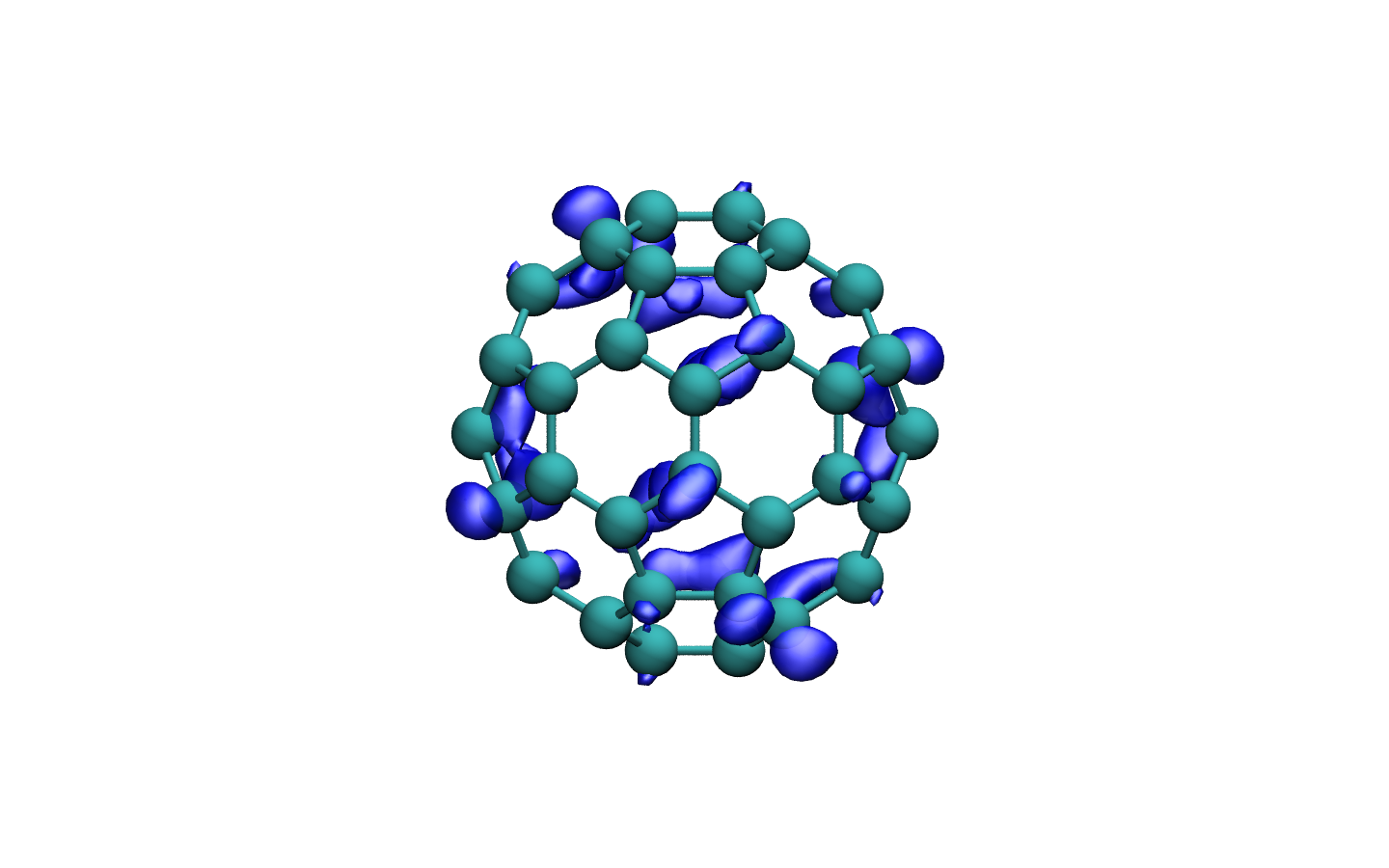}&\includegraphics[width=\wideas]{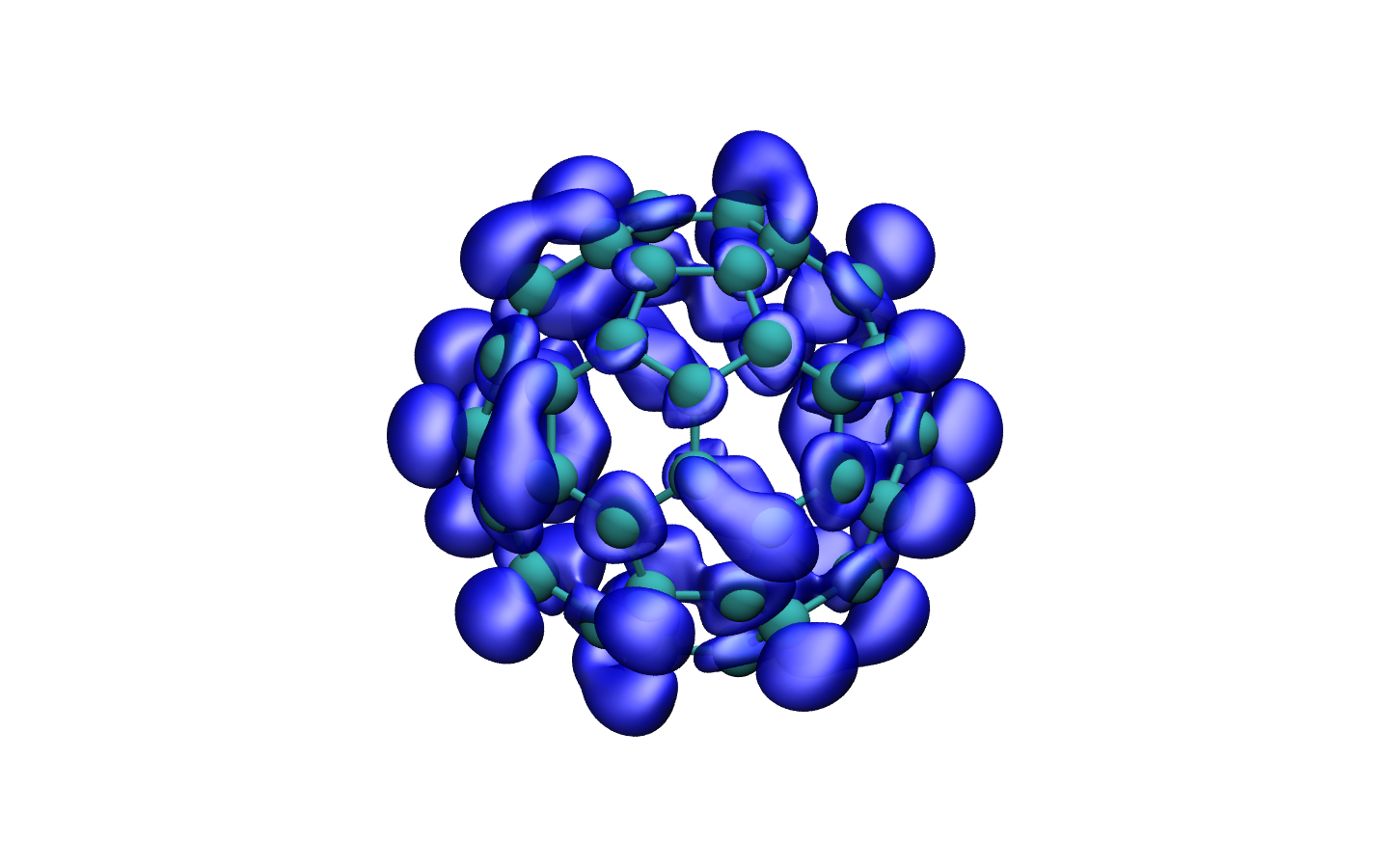}&\includegraphics[width=\wideas]{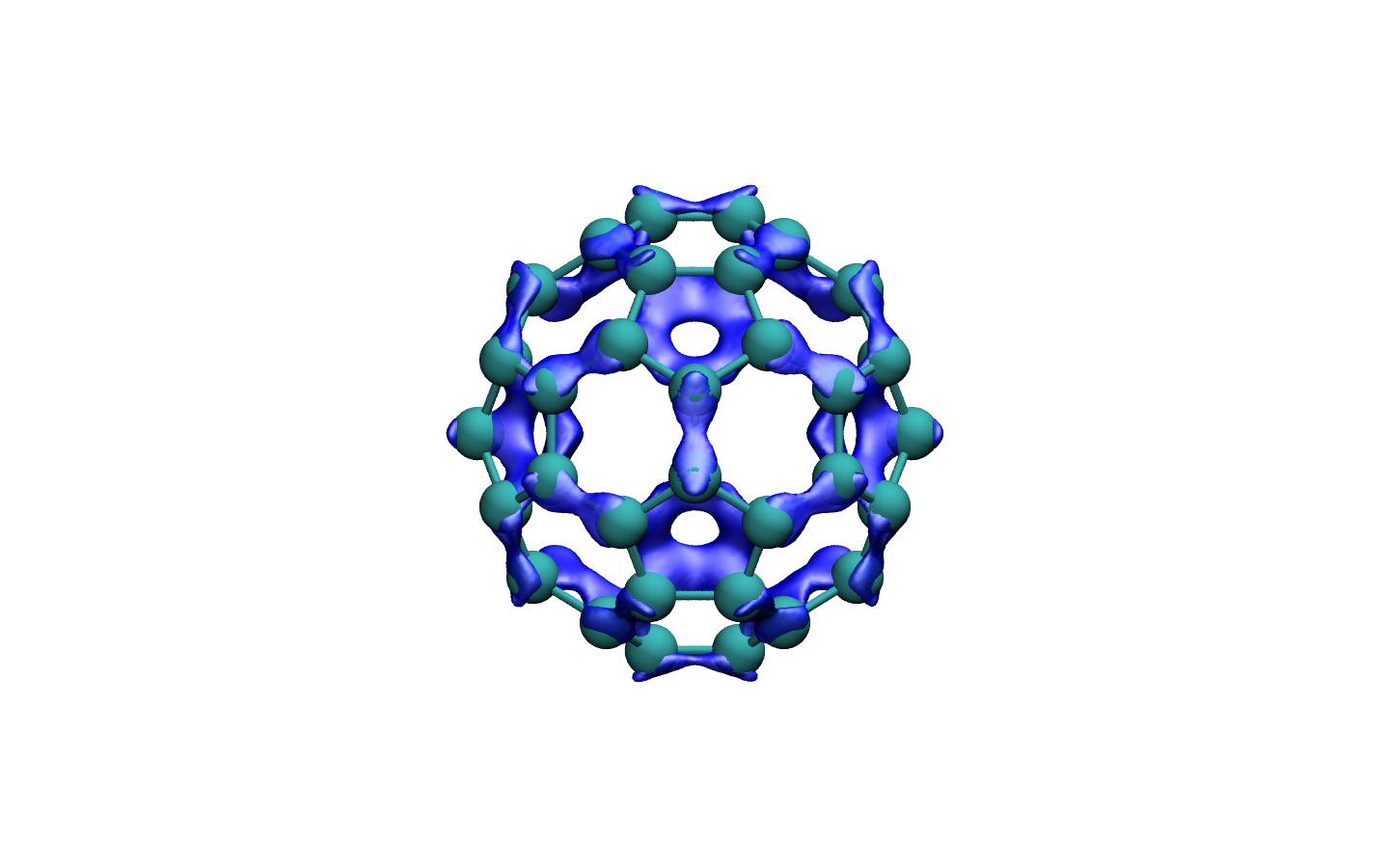}\\
	\hline
	\includegraphics[width=\wideas]{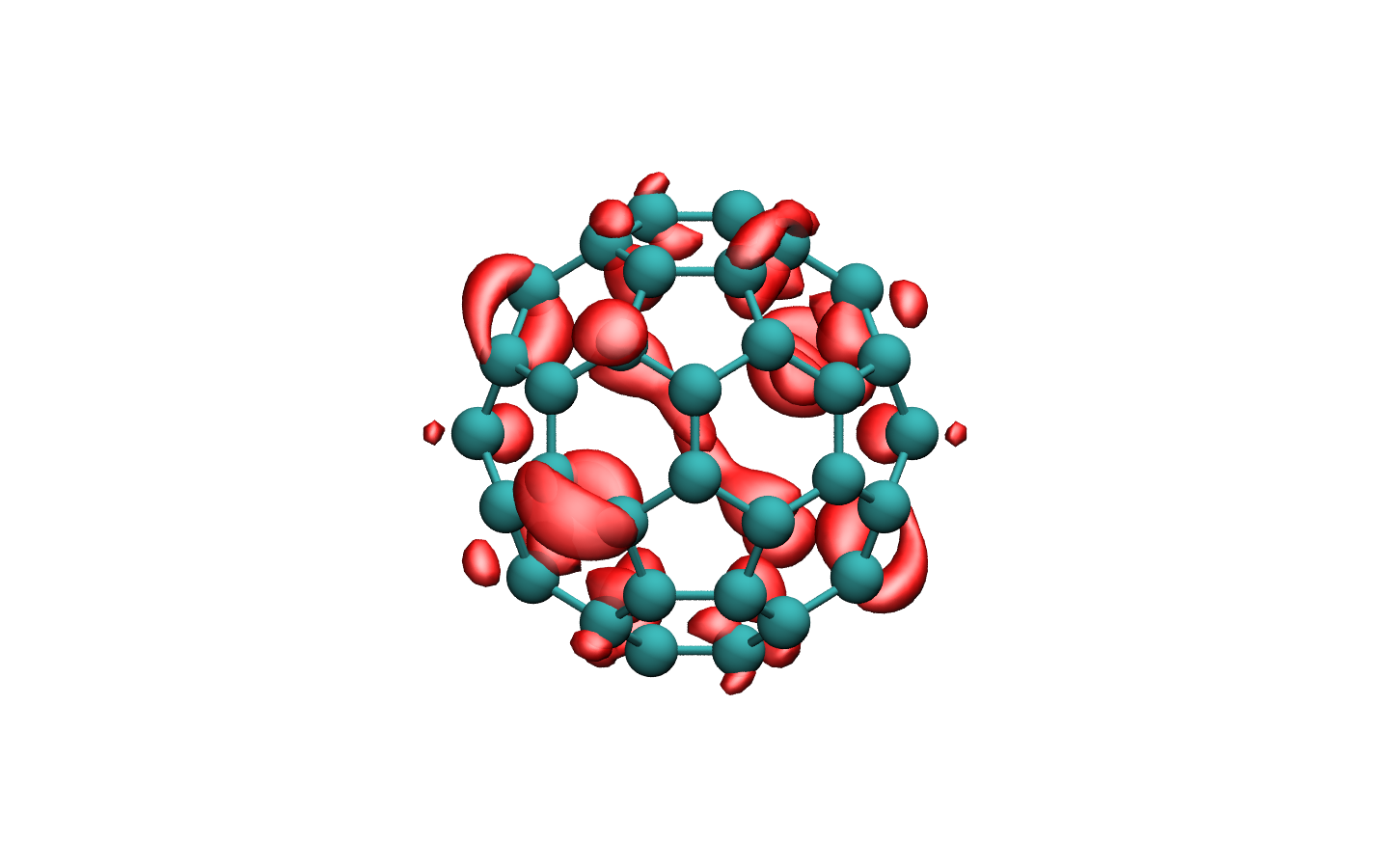}&\includegraphics[width=\wideas]{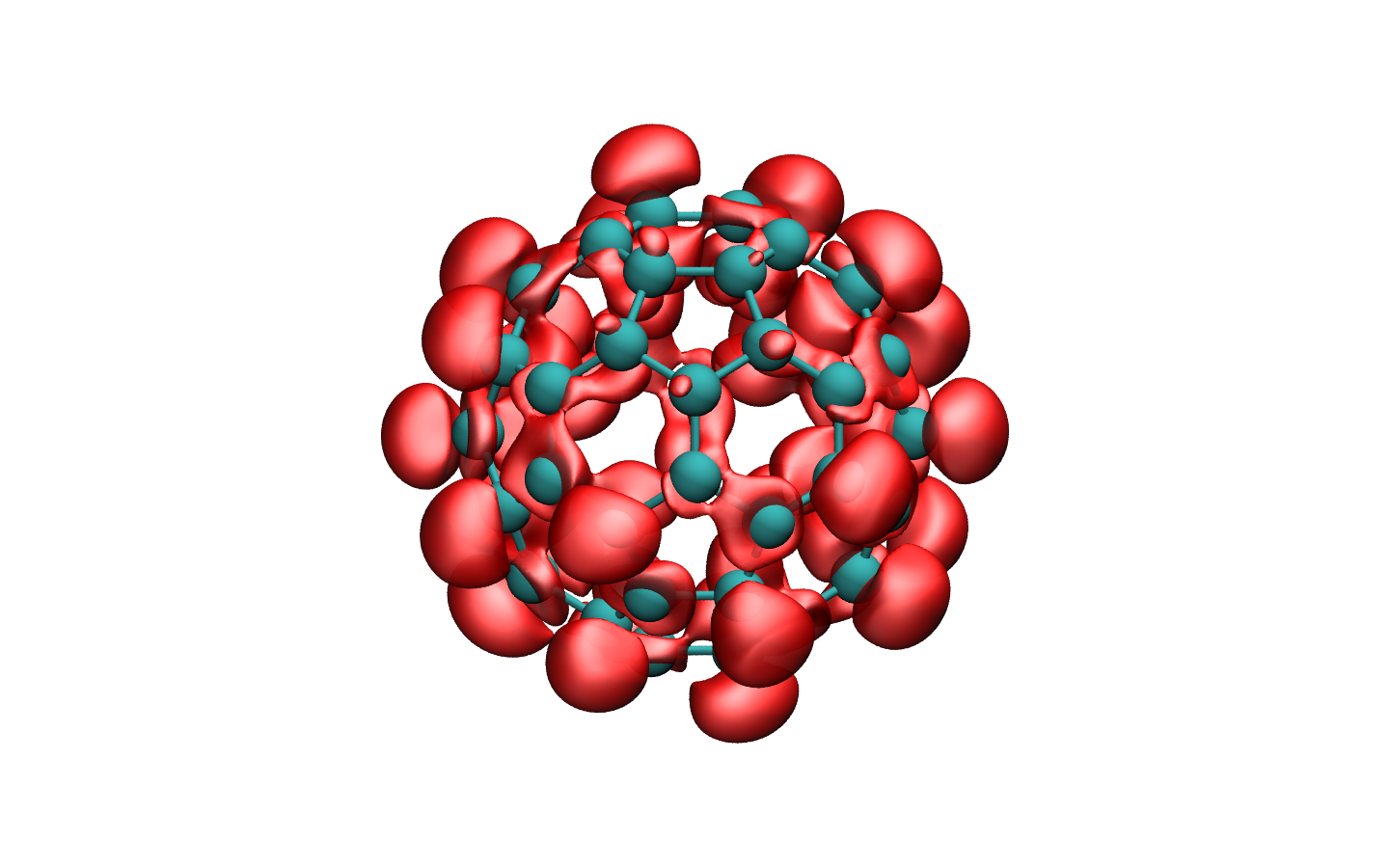}&\includegraphics[width=\wideas]{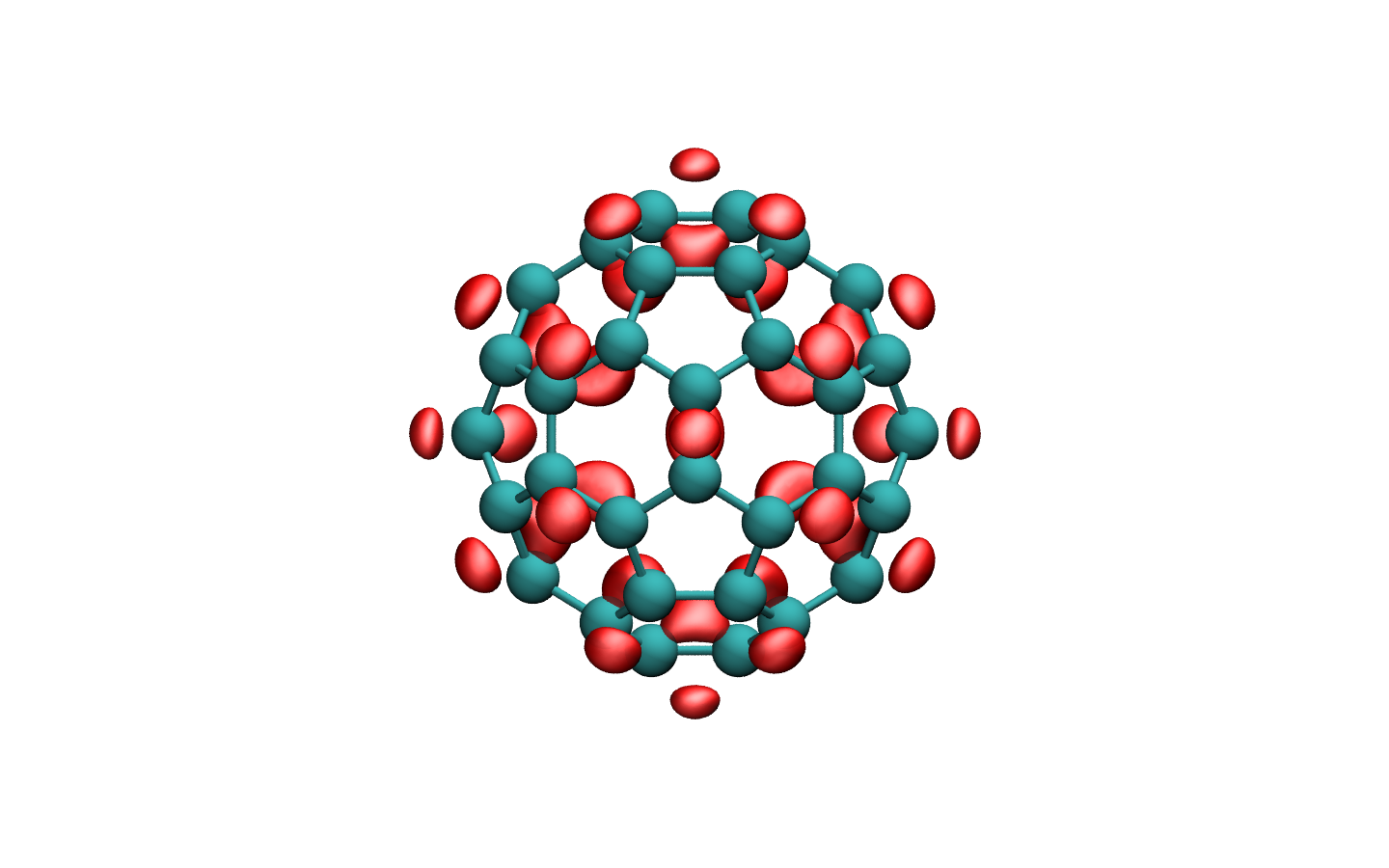}\\
\end{tabular}
}	
\end{center}
\end{figure}	

EOM-CCSD is accurate for these systems, as both single and double excitations are accounted for in the method. In Figure \ref{DDeom}, we can see that the \txcdft\ densities are improved with respect to the other methods (EOM-CCSD is not provided for fullerene due to the computational expense involved). In our previous work \cite{ramo2018a}, we showed that XCDFT densities (especially benzene) are as inaccurate as the ones computed with \dscf\, and the results of Figure \ref{DDeom} confirm this observation also for other molecules featuring exact degeneracy (such as fullerene). As expected, when $\tau > 0$, \txcdft\ is able to capture much more accurately the electron density of the excited state, while TDDFT lacks some aspects in comparison to the EOM-CCSD densities (see negative equatorial component in the EOM-CCSD density of benzene which is partially present in \txcdft\ but completely absent in TDDFT as we could verify by inflating the isosurfaces and double checking the cube files). Such nodal structure in the differential densities are commonly found in the literature \cite{saha2006,est2016} and are expected when substantial orbital relaxation occurs. 

For acrolein, there are no degeneracies and thus \txcdft\ and XCDFT deliver the same result which compare more favorably EOM-CCSD than TDDFT. 

The orbital relaxation is also seen in fullerene (see Figure \ref{full}), where the \txcdft\ differential density follows the TDDFT one but is more delocalized indicating relaxation. Unfortunately, due to the large computational expense involved, we do not have an EOM-CCSD calculation available to further confirm our findings.

From the above analysis, it is clear that the restriction in XCDFT and \dscf\ to a single Slater determinant is detrimental to the quality of the electronic structure of multireference excited states. In particular, focussing on the benzene molecule, we notice that in order to satisfy the criterion of excitation of a single electron, XCDFT's excited state orbitals are mixed ($e$ and $g$ superscripts indicate excited and ground state, respectively):
\begin{eqnarray}
\label{homoXCDFT}
\phi_{H}^{e}(\br) &=& \frac{1}{\sqrt{2}}\left[\phi_{H}^g + \phi_{L+1}^g\right],\\
\label{homo-1XCDFT}
\phi_{H-1}^{e}(\br) &=& \frac{1}{\sqrt{2}}\left[\phi_{H-1}^g + \phi_{L}^g\right].
\end{eqnarray}

As a result, the excited state wavefunction can be represented by the following superposition of configuration state functions built from the reference ground state and associated excited Slater determinants. Namely, 
\eqtn{csfXCDFT}{
\Psi_e = \frac{1}{2}\Psi_g + \frac{1}{2} \Psi_{H-1}^{L} + \frac{1}{2} \Psi_{H}^{L+1} + \frac{1}{2} \Psi_{H-1,H}^{L,L+1}.
}
The above, clearly indicates that the XCDFT excited state wavefunction, $\Psi_e$, has strong overlap with the ground state wavefunction, $\Psi_g$, and an equally strong double excitation character arising from the $\Psi_{H-1,H}^{L,L+1}$ term.

In our trial calculations (not reported), we have noticed that the above described issue is shared among aromatic chromophores, casting serious doubts about the physicality of \dscf\ excited states which are frequently used as initial conditions for nonadiabatic dynamics simulations.
\def\wideas{0.25\textwidth}
\begin{figure}[htp]
\caption{\label{benzorb} Comparison of the frontier molecular orbitals of benzene obtained with \txcdft\ and XCDFT against the corresponding ground state orbitals. The occupation numbers of these orbitals are shown. }
\begin{center}
{\fontsize{16}{18}\selectfont
\begin{tabular}{ccc}
	Ground State & XCDFT & $\tau$--XCDFT\\
	\hline
	1&1&0.5\\
	\includegraphics[width=\wideas]{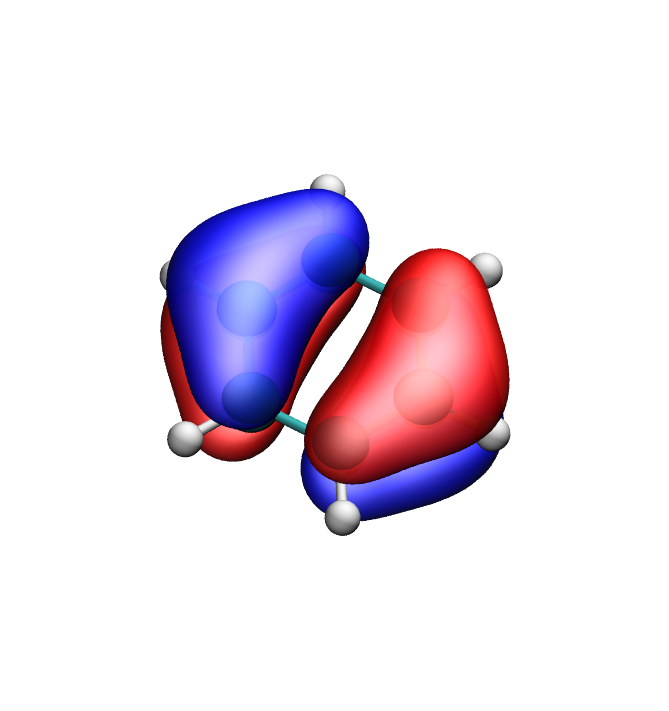}&\includegraphics[width=\wideas]{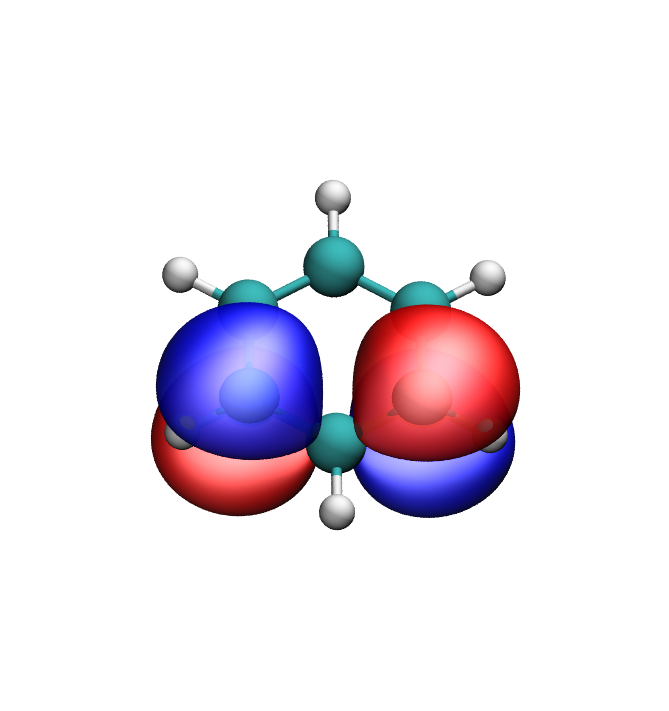}&\includegraphics[width=\wideas]{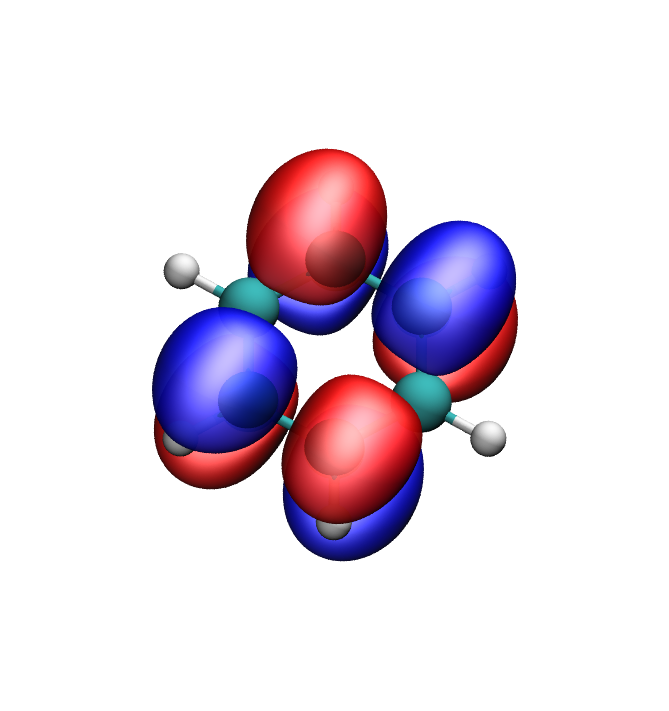}\\
	\hline
	1&1&0.5\\
	\includegraphics[width=\wideas]{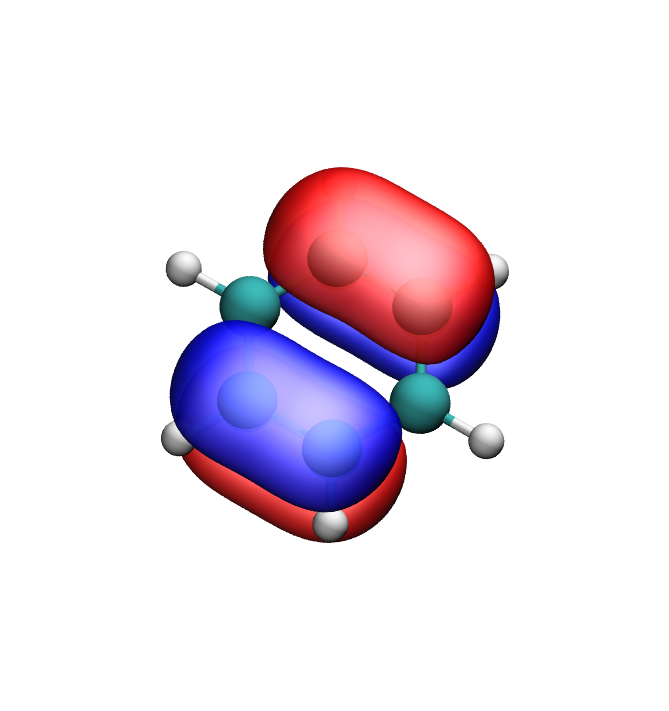}&\includegraphics[width=\wideas]{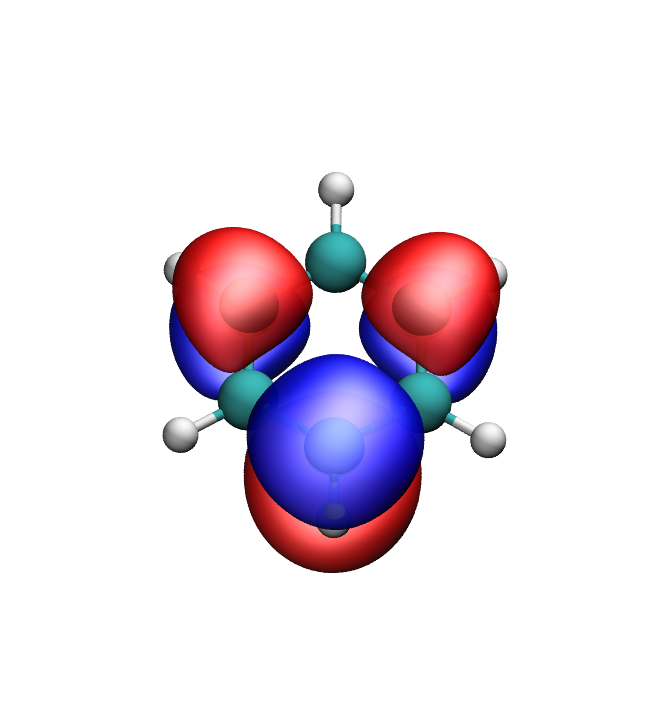}&\includegraphics[width=\wideas]{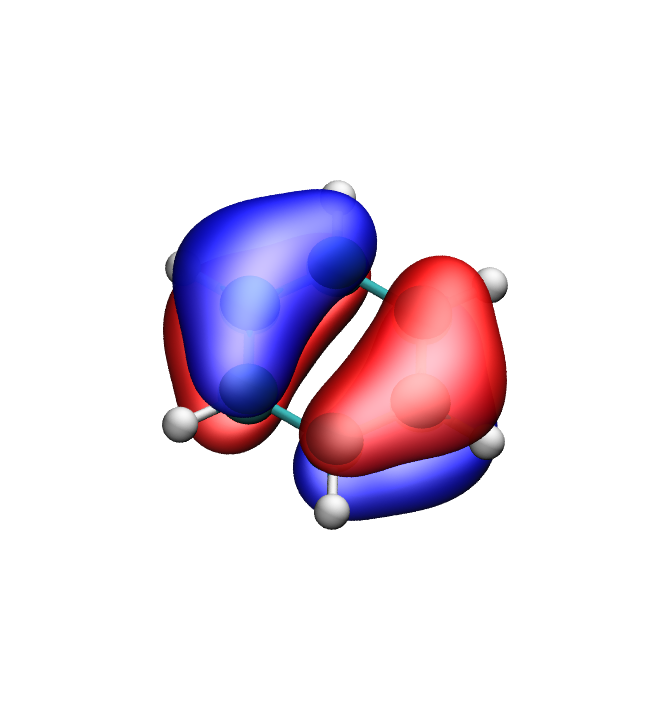}\\
	\hline
	0&0&0.5\\
	\includegraphics[width=\wideas]{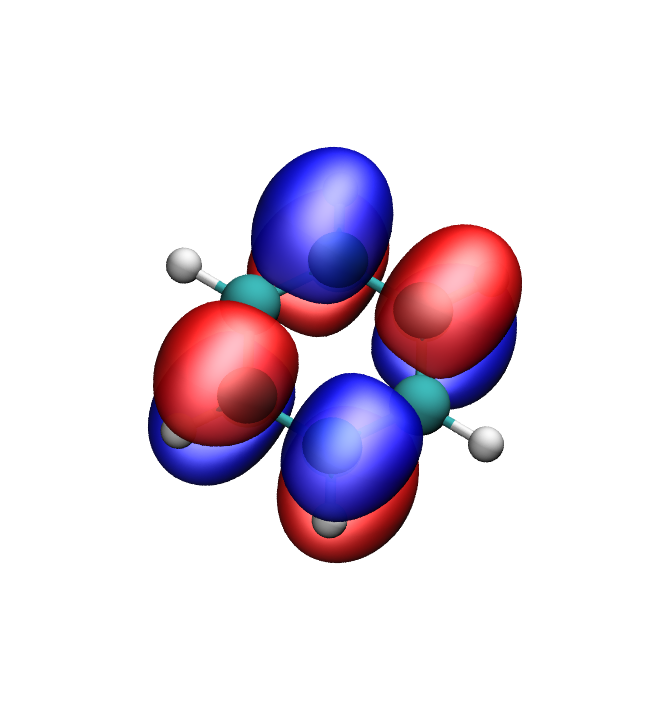}&\includegraphics[width=\wideas]{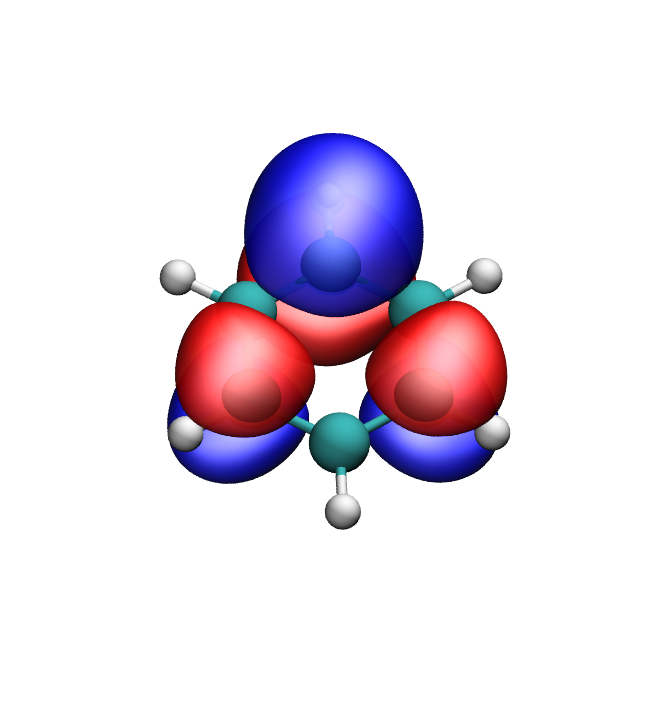}&\includegraphics[width=\wideas]{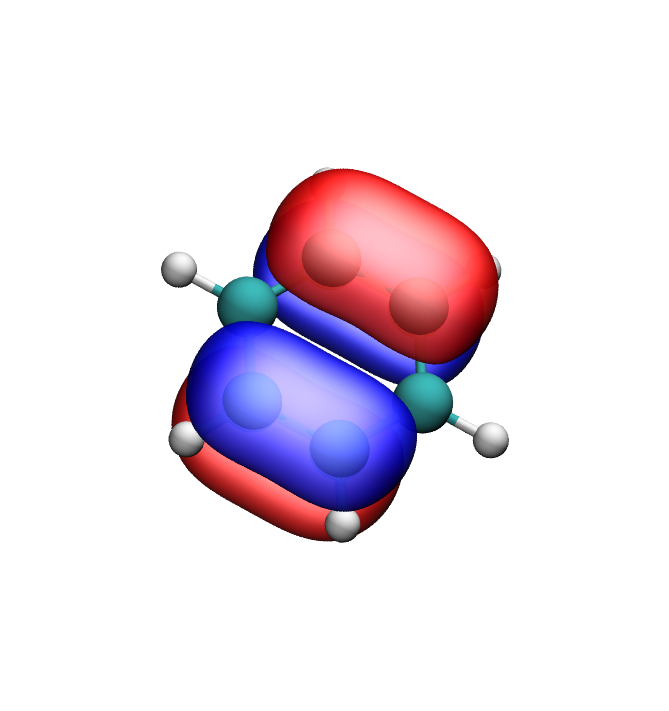}\\
	\hline
	0&0&0.5\\
	\includegraphics[width=\wideas]{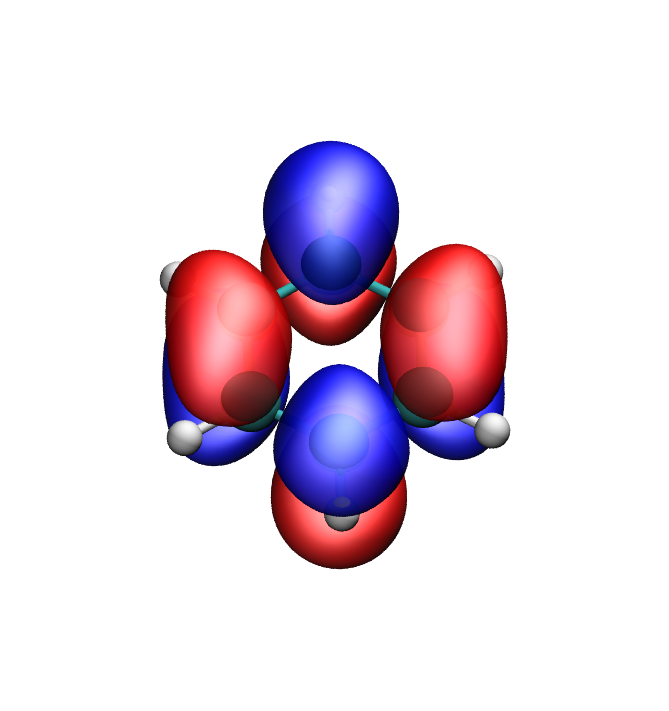}&\includegraphics[width=\wideas]{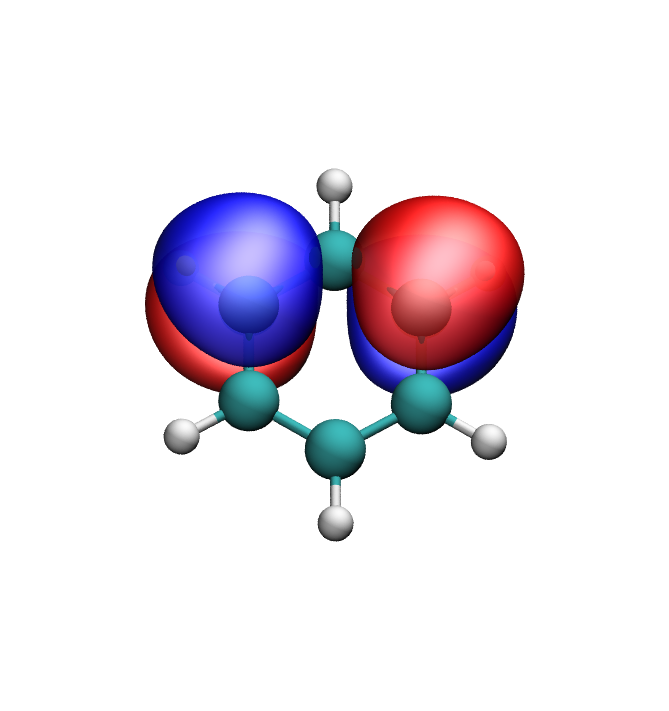}&\includegraphics[width=\wideas]{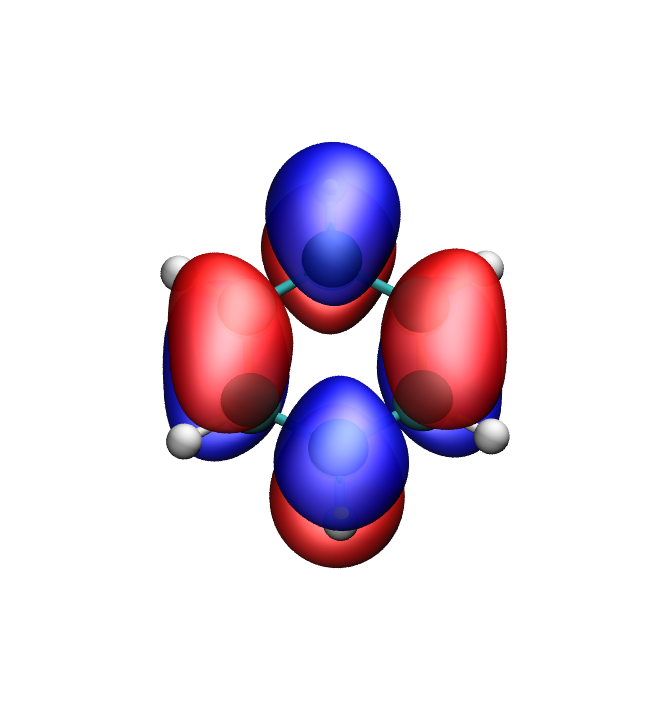}\\
\end{tabular}	
}
\end{center} 	
\end{figure}

In Figure \ref{benzorb}, we plot the frontier occupied and virtual orbitals of the ground, XCDFT and \txcdft\ excited states. The figure indicates that the \txcdft\ orbitals largely resemble the ground state orbitals with some small deviations due to orbital relaxation effects (typically accounted for in wavefunction methods by high order excitation contributions). Instead, the XCDFT orbitals are very different from the ground state ones indicating that in XCDFT, in order to satisfy the imposed constraint in \eqn{vc}, the frontier orbitals have mixed and rotated dramatically exposing an unphysical character. Thus, we can conclude this analysis by stating that for multireference excited states, \txcdft\ orbitals indicate a degree of relaxation compared to the ground state orbitals while still retaining the overall character resulting in differential densities in agreement with EOM-CCSD calculations. A similar analysis can be carried out for Fullerene, although it is not reported.

\subsection{Quality of the excitation energies}
We summarize in Table \ref{tab1} excitation energies computed with \txcdft\ along side with available benchmark data \cite{Schrei2008a,Schrei2008b,Martins2009,Holland2014,Floris2014,Zimmer2010}. We find that the performance of \txcdft\ is comparable with XCDFT \cite{ramo2018a}. However, as we have analyzed above, the character of the states involved are now corrected. 

\begin{table}[htp]
	\caption{\label{tab1} \txcdft\ excitation energy values (in eV) for all exchange-correlation functionals considered.}	
\begin{center}
\begin{tabular}{lccccc}
\Xhline{3\arrayrulewidth}
	System  & PBE & M06--L& SCAN & revTPSS & Benchmark\\ 
\hline	
	Ethylene &6.06 & 6.14 &  5.90 &  6.08       & 7.80\\
	Tetrafluoroethylene &6.23 &6.59 & 6.42 & 6.35&7.08\\ 
	Isoprene &4.99 &4.53 & 4.35 & 5.32           &5.74\\ 
	1,3-Butadiene &4.56 &4.51 & 4.35 & 4.56      &6.18\\ 
	Formaldehyde  &3.95 &3.48 & 2.86 & 3.48      &3.88\\ 
	Propanamide   &5.76 &5.51 & 5.11 & 5.79      &5.72\\ 
	Acrolein     &3.89 &3.28 & 2.68 & 3.42      &3.75\\ 
	Pyrrole       &5.46 &5.73 & 5.52 & 5.57      &6.37\\ 
	Thiophene     &5.28 &5.30 & 5.03 & 5.13      &5.64\\ 
	Benzaldehyde  &3.75 &3.31 & 2.67 & 3.49      &3.34\\ 
	Adenine &4.55 &4.68 & 4.46 & 4.66            &5.25\\ 
	Cytosine &4.31 &4.87 & 4.67 & 4.61           &4.66\\ 
	Benzene	 &5.19 &5.47 & 5.37 & 5.36           &5.08\\ 
	Naphthalene &4.01 &3.94 & 3.77 & 3.97        &4.24\\ 
	Anthracene &3.12 &3.04 & 2.89 & 3.06         &3.55\\ 
	Tetracene &2.14 &2.03 & 1.90 & 2.06          &2.95\\ 
	Pentacene &1.96 &1.87 & 1.73 & 1.88          &2.30\\ 
	Fullerene &1.59 &1.73 & 1.69 & 1.65          &1.75 \\
\Xhline{3\arrayrulewidth}
\end{tabular}

\end{center}
\end{table}

\begin{figure}
 \includegraphics[width=1.0\textwidth]{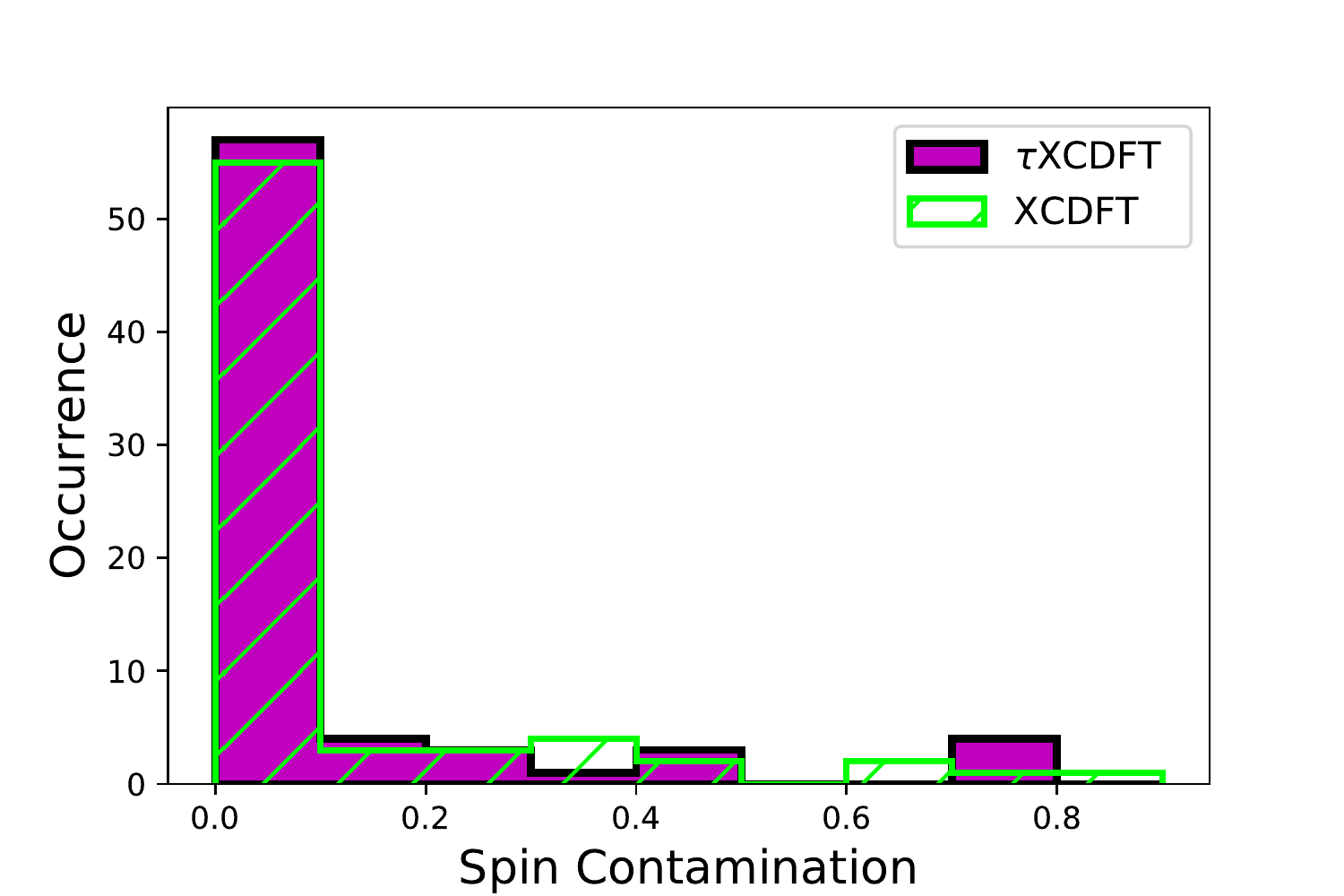}
        \caption{Histogram of the spin contamination for all XCDFT and \txcdft\ excited states collecting all exchange-correlation functionals.}
\label{sp}
\end{figure}

In Table \ref{tab2}, we show the mean unsigned error (MUE) for the excitation energies computed with \txcdft, XCDFT, \dscf, and TDDFT. The MUE shows that \txcdft\ and XCDFT are comparable to TDDFT and significantly better than \dscf. We see that among all metaGGA functionals, revTPSS is the better performing. In an effort to explain some of the trends, in Figure \ref{sp} we report a histogram of a measure of spin contamination \txcdft\ and XCDFT collecting all exchange-correlation functionanls considered. The histograms show that overall the spin contamination is well handled by XCDFT and \txcdft. However, we notice that in \txcdft\ the spin contamination is less prevalent, and we also note that the systems with high contamination (above 0.5) correspond to benzene and fullerene (i.e., where there are strong degeneracies among the frontier orbitals) computed with the SCAN functional.

\begin{table}[htp]
        \caption{\label{tab2} Mean unsigned error (MUE) against benchmark values across the entire set of all excitation energies computed for all exchange-correlation functionals and methods considered.}
\begin{center}
\begin{tabular}{lcccc}
\Xhline{3\arrayrulewidth}
        Method  & PBE & M06--L & SCAN & revTPSS \\
\hline
        XCDFT & 0.677 & 0.447 & 0.378 & 0.503\\
        \txcdft\ & 0.566 & 0.591 & 0.816 & 0.557\\
        $\Delta$SCF & 1.320 & 0.790 & 1.660 & 1.420\\
        TDDFT & 0.390 & 0.620 & 0.513 & 0.375\\
\Xhline{3\arrayrulewidth}
\end{tabular}

\end{center}
\end{table}

\section{Conclusions}
In conclusion, we developed, implemented in the ADF program, and benchmarked a mean field method for the computation of low-lying electronic excited states, \txcdft. This method is capable of accounting for degenerate energy levels often present in excited states, such as aromatic chromophores. We show that quality low-lying excited states are found by using ensemble 1-RDMs. We also show that when considering multireference excited states, mean field methods that employ a single Slater determinant (such as \dscf\ and XCDFT) completely fail in predicting the electronic structure. \txcdft, instead reproduces the electronic density of these excited states, avoids incorrect rotation among the frontier orbitals and correctly features effects of orbital relaxation. 

\begin{acknowledgments} 
This material is based upon work supported by the U.S. Department of Energy, Office of Basic Energy Sciences, under Award Number DE-SC0018343. The authors acknowledge the Office of Advanced Research Computing (OARC) at Rutgers, The State University of New Jersey for providing access to the Amarel cluster and associated research computing resources that have contributed to the results reported here. URL: http://oarc.rutgers.edu 
\end{acknowledgments}
\bibliography{prg_bibliography/prg}

\begin{thebibliography}{41}%
\makeatletter
\providecommand \@ifxundefined [1]{%
 \@ifx{#1\undefined}
}%
\providecommand \@ifnum [1]{%
 \ifnum #1\expandafter \@firstoftwo
 \else \expandafter \@secondoftwo
 \fi
}%
\providecommand \@ifx [1]{%
 \ifx #1\expandafter \@firstoftwo
 \else \expandafter \@secondoftwo
 \fi
}%
\providecommand \natexlab [1]{#1}%
\providecommand \enquote  [1]{``#1''}%
\providecommand \bibnamefont  [1]{#1}%
\providecommand \bibfnamefont [1]{#1}%
\providecommand \citenamefont [1]{#1}%
\providecommand \href@noop [0]{\@secondoftwo}%
\providecommand \href [0]{\begingroup \@sanitize@url \@href}%
\providecommand \@href[1]{\@@startlink{#1}\@@href}%
\providecommand \@@href[1]{\endgroup#1\@@endlink}%
\providecommand \@sanitize@url [0]{\catcode `\\12\catcode `\$12\catcode
  `\&12\catcode `\#12\catcode `\^12\catcode `\_12\catcode `\%12\relax}%
\providecommand \@@startlink[1]{}%
\providecommand \@@endlink[0]{}%
\providecommand \url  [0]{\begingroup\@sanitize@url \@url }%
\providecommand \@url [1]{\endgroup\@href {#1}{\urlprefix }}%
\providecommand \urlprefix  [0]{URL }%
\providecommand \Eprint [0]{\href }%
\providecommand \doibase [0]{http://dx.doi.org/}%
\providecommand \selectlanguage [0]{\@gobble}%
\providecommand \bibinfo  [0]{\@secondoftwo}%
\providecommand \bibfield  [0]{\@secondoftwo}%
\providecommand \translation [1]{[#1]}%
\providecommand \BibitemOpen [0]{}%
\providecommand \bibitemStop [0]{}%
\providecommand \bibitemNoStop [0]{.\EOS\space}%
\providecommand \EOS [0]{\spacefactor3000\relax}%
\providecommand \BibitemShut  [1]{\csname bibitem#1\endcsname}%
\let\auto@bib@innerbib\@empty
\bibitem [{\citenamefont {Filatov}(2014)}]{Filatov_2014}%
  \BibitemOpen
  \bibfield  {author} {\bibinfo {author} {\bibfnamefont {M.}~\bibnamefont
  {Filatov}},\ }\href {\doibase 10.1002/wcms.1209} {\bibfield  {journal}
  {\bibinfo  {journal} {WIREs: Comput. Mol. Sci.}\ }\textbf {\bibinfo {volume}
  {5}},\ \bibinfo {pages} {146} (\bibinfo {year} {2014})}\BibitemShut {NoStop}%
\bibitem [{\citenamefont {Filatov}\ \emph {et~al.}(2017)\citenamefont
  {Filatov}, \citenamefont {Mart{\'{\i}}nez},\ and\ \citenamefont
  {Kim}}]{Filatov_2017}%
  \BibitemOpen
  \bibfield  {author} {\bibinfo {author} {\bibfnamefont {M.}~\bibnamefont
  {Filatov}}, \bibinfo {author} {\bibfnamefont {T.~J.}\ \bibnamefont
  {Mart{\'{\i}}nez}}, \ and\ \bibinfo {author} {\bibfnamefont {K.~S.}\
  \bibnamefont {Kim}},\ }\href {\doibase 10.1063/1.4996873} {\bibfield
  {journal} {\bibinfo  {journal} {{J. Chem. Phys.}}\ }\textbf {\bibinfo
  {volume} {147}},\ \bibinfo {pages} {064104} (\bibinfo {year}
  {2017})}\BibitemShut {NoStop}%
\bibitem [{\citenamefont {hui Yang}\ \emph {et~al.}(2017)\citenamefont {hui
  Yang}, \citenamefont {Pribram-Jones}, \citenamefont {Burke},\ and\
  \citenamefont {Ullrich}}]{Yang_2017}%
  \BibitemOpen
  \bibfield  {author} {\bibinfo {author} {\bibfnamefont {Z.}~\bibnamefont {hui
  Yang}}, \bibinfo {author} {\bibfnamefont {A.}~\bibnamefont {Pribram-Jones}},
  \bibinfo {author} {\bibfnamefont {K.}~\bibnamefont {Burke}}, \ and\ \bibinfo
  {author} {\bibfnamefont {C.~A.}\ \bibnamefont {Ullrich}},\ }\href {\doibase
  10.1103/physrevlett.119.033003} {\bibfield  {journal} {\bibinfo  {journal}
  {{Phys. Rev. Lett.}}\ }\textbf {\bibinfo {volume} {119}} (\bibinfo {year}
  {2017}),\ 10.1103/physrevlett.119.033003}\BibitemShut {NoStop}%
\bibitem [{\citenamefont {Deur}\ and\ \citenamefont
  {Fromager}(2019)}]{Deur_2019}%
  \BibitemOpen
  \bibfield  {author} {\bibinfo {author} {\bibfnamefont {K.}~\bibnamefont
  {Deur}}\ and\ \bibinfo {author} {\bibfnamefont {E.}~\bibnamefont
  {Fromager}},\ }\href {\doibase 10.1063/1.5084312} {\bibfield  {journal}
  {\bibinfo  {journal} {{J. Chem. Phys.}}\ }\textbf {\bibinfo {volume} {150}},\
  \bibinfo {pages} {094106} (\bibinfo {year} {2019})}\BibitemShut {NoStop}%
\bibitem [{\citenamefont {Franck}\ and\ \citenamefont
  {Fromager}(2013)}]{Franck_2013}%
  \BibitemOpen
  \bibfield  {author} {\bibinfo {author} {\bibfnamefont {O.}~\bibnamefont
  {Franck}}\ and\ \bibinfo {author} {\bibfnamefont {E.}~\bibnamefont
  {Fromager}},\ }\href {\doibase 10.1080/00268976.2013.858191} {\bibfield
  {journal} {\bibinfo  {journal} {{Mol. Phys.}}\ }\textbf {\bibinfo {volume}
  {112}},\ \bibinfo {pages} {1684} (\bibinfo {year} {2013})}\BibitemShut
  {NoStop}%
\bibitem [{\citenamefont {Evangelista}\ \emph {et~al.}(2013)\citenamefont
  {Evangelista}, \citenamefont {Shushkov},\ and\ \citenamefont
  {Tully}}]{evan2013}%
  \BibitemOpen
  \bibfield  {author} {\bibinfo {author} {\bibfnamefont {F.~A.}\ \bibnamefont
  {Evangelista}}, \bibinfo {author} {\bibfnamefont {P.}~\bibnamefont
  {Shushkov}}, \ and\ \bibinfo {author} {\bibfnamefont {J.~C.}\ \bibnamefont
  {Tully}},\ }\href {\doibase 10.1021/jp401323d} {\bibfield  {journal}
  {\bibinfo  {journal} {{J. Phys. Chem. A}}\ }\textbf {\bibinfo {volume}
  {117}},\ \bibinfo {pages} {7378} (\bibinfo {year} {2013})}\BibitemShut
  {NoStop}%
\bibitem [{\citenamefont {Ramos}\ and\ \citenamefont
  {Pavanello}(2018)}]{ramo2018a}%
  \BibitemOpen
  \bibfield  {author} {\bibinfo {author} {\bibfnamefont {P.}~\bibnamefont
  {Ramos}}\ and\ \bibinfo {author} {\bibfnamefont {M.}~\bibnamefont
  {Pavanello}},\ }\href {\doibase 10.1063/1.5018615} {\bibfield  {journal}
  {\bibinfo  {journal} {{J. Chem. Phys.}}\ }\textbf {\bibinfo {volume} {148}},\
  \bibinfo {pages} {144103} (\bibinfo {year} {2018})},\ \Eprint
  {http://arxiv.org/abs/https://doi.org/10.1063/1.5018615}
  {https://doi.org/10.1063/1.5018615} \BibitemShut {NoStop}%
\bibitem [{\citenamefont {Kaduk}\ \emph {et~al.}(2012)\citenamefont {Kaduk},
  \citenamefont {Kowalczyk},\ and\ \citenamefont {{Van Voorhis}}}]{kadu2012}%
  \BibitemOpen
  \bibfield  {author} {\bibinfo {author} {\bibfnamefont {B.}~\bibnamefont
  {Kaduk}}, \bibinfo {author} {\bibfnamefont {T.}~\bibnamefont {Kowalczyk}}, \
  and\ \bibinfo {author} {\bibfnamefont {T.}~\bibnamefont {{Van Voorhis}}},\
  }\href {\doibase 10.1021/cr200148b} {\bibfield  {journal} {\bibinfo
  {journal} {{Chem. Rev.}}\ }\textbf {\bibinfo {volume} {112}},\ \bibinfo
  {pages} {321} (\bibinfo {year} {2012})},\ \Eprint
  {http://arxiv.org/abs/http://pubs.acs.org/doi/pdf/10.1021/cr200148b}
  {http://pubs.acs.org/doi/pdf/10.1021/cr200148b} \BibitemShut {NoStop}%
\bibitem [{\citenamefont {Cembran}\ \emph {et~al.}(2010)\citenamefont
  {Cembran}, \citenamefont {Payaka}, \citenamefont {lin Lin}, \citenamefont
  {Xie}, \citenamefont {Mo}, \citenamefont {Song},\ and\ \citenamefont
  {Gao}}]{Cembran_2010}%
  \BibitemOpen
  \bibfield  {author} {\bibinfo {author} {\bibfnamefont {A.}~\bibnamefont
  {Cembran}}, \bibinfo {author} {\bibfnamefont {A.}~\bibnamefont {Payaka}},
  \bibinfo {author} {\bibfnamefont {Y.}~\bibnamefont {lin Lin}}, \bibinfo
  {author} {\bibfnamefont {W.}~\bibnamefont {Xie}}, \bibinfo {author}
  {\bibfnamefont {Y.}~\bibnamefont {Mo}}, \bibinfo {author} {\bibfnamefont
  {L.}~\bibnamefont {Song}}, \ and\ \bibinfo {author} {\bibfnamefont
  {J.}~\bibnamefont {Gao}},\ }\href {\doibase 10.1021/ct1001686} {\bibfield
  {journal} {\bibinfo  {journal} {{J. Chem. Theory Comput.}}\ }\textbf
  {\bibinfo {volume} {6}},\ \bibinfo {pages} {2242} (\bibinfo {year}
  {2010})}\BibitemShut {NoStop}%
\bibitem [{\citenamefont {Cembran}\ \emph {et~al.}(2009)\citenamefont
  {Cembran}, \citenamefont {Song}, \citenamefont {Mo},\ and\ \citenamefont
  {Gao}}]{cem2009}%
  \BibitemOpen
  \bibfield  {author} {\bibinfo {author} {\bibfnamefont {A.}~\bibnamefont
  {Cembran}}, \bibinfo {author} {\bibfnamefont {L.}~\bibnamefont {Song}},
  \bibinfo {author} {\bibfnamefont {Y.}~\bibnamefont {Mo}}, \ and\ \bibinfo
  {author} {\bibfnamefont {J.}~\bibnamefont {Gao}},\ }\href@noop {} {\bibfield
  {journal} {\bibinfo  {journal} {{J. Chem. Theory Comput.}}\ }\textbf
  {\bibinfo {volume} {5}},\ \bibinfo {pages} {2702} (\bibinfo {year}
  {2009})}\BibitemShut {NoStop}%
\bibitem [{\citenamefont {Grimme}\ and\ \citenamefont
  {Waletzke}(1999)}]{Grimme_1999}%
  \BibitemOpen
  \bibfield  {author} {\bibinfo {author} {\bibfnamefont {S.}~\bibnamefont
  {Grimme}}\ and\ \bibinfo {author} {\bibfnamefont {M.}~\bibnamefont
  {Waletzke}},\ }\href {\doibase 10.1063/1.479866} {\bibfield  {journal}
  {\bibinfo  {journal} {{J. Chem. Phys.}}\ }\textbf {\bibinfo {volume} {111}},\
  \bibinfo {pages} {5645} (\bibinfo {year} {1999})}\BibitemShut {NoStop}%
\bibitem [{\citenamefont {Mei}\ and\ \citenamefont {Yang}(2019)}]{Mei_2019}%
  \BibitemOpen
  \bibfield  {author} {\bibinfo {author} {\bibfnamefont {Y.}~\bibnamefont
  {Mei}}\ and\ \bibinfo {author} {\bibfnamefont {W.}~\bibnamefont {Yang}},\
  }\href {\doibase 10.1021/acs.jpclett.9b00712} {\bibfield  {journal} {\bibinfo
   {journal} {{J. Phys. Chem. Lett.}}\ }\textbf {\bibinfo {volume} {10}},\
  \bibinfo {pages} {2538} (\bibinfo {year} {2019})}\BibitemShut {NoStop}%
\bibitem [{\citenamefont {{Van Voorhis}}\ \emph {et~al.}(2010)\citenamefont
  {{Van Voorhis}}, \citenamefont {Kowalczyk}, \citenamefont {Kaduk},
  \citenamefont {Wang}, \citenamefont {Cheng},\ and\ \citenamefont
  {Wu}}]{vanv2010a}%
  \BibitemOpen
  \bibfield  {author} {\bibinfo {author} {\bibfnamefont {T.}~\bibnamefont {{Van
  Voorhis}}}, \bibinfo {author} {\bibfnamefont {T.}~\bibnamefont {Kowalczyk}},
  \bibinfo {author} {\bibfnamefont {B.}~\bibnamefont {Kaduk}}, \bibinfo
  {author} {\bibfnamefont {L.-P.}\ \bibnamefont {Wang}}, \bibinfo {author}
  {\bibfnamefont {C.-L.}\ \bibnamefont {Cheng}}, \ and\ \bibinfo {author}
  {\bibfnamefont {Q.}~\bibnamefont {Wu}},\ }\href@noop {} {\bibfield  {journal}
  {\bibinfo  {journal} {Annu. Rev. Phys. Chem.}\ }\textbf {\bibinfo {volume}
  {61}},\ \bibinfo {pages} {149} (\bibinfo {year} {2010})}\BibitemShut
  {NoStop}%
\bibitem [{\citenamefont {Wu}\ and\ \citenamefont {{Van
  Voorhis}}(2006)}]{wu2006}%
  \BibitemOpen
  \bibfield  {author} {\bibinfo {author} {\bibfnamefont {Q.}~\bibnamefont
  {Wu}}\ and\ \bibinfo {author} {\bibfnamefont {T.}~\bibnamefont {{Van
  Voorhis}}},\ }\href@noop {} {\bibfield  {journal} {\bibinfo  {journal} {{J.
  Chem. Phys.}}\ }\textbf {\bibinfo {volume} {125}},\ \bibinfo {pages} {164105}
  (\bibinfo {year} {2006})}\BibitemShut {NoStop}%
\bibitem [{\citenamefont {Park}\ \emph {et~al.}(2016)\citenamefont {Park},
  \citenamefont {Senn}, \citenamefont {Krykunov},\ and\ \citenamefont
  {Ziegler}}]{zieg2016}%
  \BibitemOpen
  \bibfield  {author} {\bibinfo {author} {\bibfnamefont {Y.~C.}\ \bibnamefont
  {Park}}, \bibinfo {author} {\bibfnamefont {F.}~\bibnamefont {Senn}}, \bibinfo
  {author} {\bibfnamefont {M.}~\bibnamefont {Krykunov}}, \ and\ \bibinfo
  {author} {\bibfnamefont {T.}~\bibnamefont {Ziegler}},\ }\href {\doibase
  10.1021/acs.jctc.6b00333} {\bibfield  {journal} {\bibinfo  {journal} {{J.
  Chem. Theory Comput.}}\ }\textbf {\bibinfo {volume} {12}},\ \bibinfo {pages}
  {5438} (\bibinfo {year} {2016})},\ \Eprint
  {http://arxiv.org/abs/http://dx.doi.org/10.1021/acs.jctc.6b00333}
  {http://dx.doi.org/10.1021/acs.jctc.6b00333} \BibitemShut {NoStop}%
\bibitem [{\citenamefont {Seidu}\ \emph {et~al.}(2014)\citenamefont {Seidu},
  \citenamefont {Krykunov},\ and\ \citenamefont {Ziegler}}]{ziegl2014}%
  \BibitemOpen
  \bibfield  {author} {\bibinfo {author} {\bibfnamefont {I.}~\bibnamefont
  {Seidu}}, \bibinfo {author} {\bibfnamefont {M.}~\bibnamefont {Krykunov}}, \
  and\ \bibinfo {author} {\bibfnamefont {T.}~\bibnamefont {Ziegler}},\ }\href
  {\doibase 10.1080/00268976.2013.852261} {\bibfield  {journal} {\bibinfo
  {journal} {{Mol. Phys.}}\ }\textbf {\bibinfo {volume} {112}},\ \bibinfo
  {pages} {661} (\bibinfo {year} {2014})},\ \Eprint
  {http://arxiv.org/abs/http://dx.doi.org/10.1080/00268976.2013.852261}
  {http://dx.doi.org/10.1080/00268976.2013.852261} \BibitemShut {NoStop}%
\bibitem [{\citenamefont {Ziegler}\ \emph {et~al.}(2012)\citenamefont
  {Ziegler}, \citenamefont {Krykunov},\ and\ \citenamefont
  {Cullen}}]{Ziegler2012}%
  \BibitemOpen
  \bibfield  {author} {\bibinfo {author} {\bibfnamefont {T.}~\bibnamefont
  {Ziegler}}, \bibinfo {author} {\bibfnamefont {M.}~\bibnamefont {Krykunov}}, \
  and\ \bibinfo {author} {\bibfnamefont {J.}~\bibnamefont {Cullen}},\ }\href
  {\doibase 10.1063/1.3696967} {\bibfield  {journal} {\bibinfo  {journal} {{J.
  Chem. Phys.}}\ }\textbf {\bibinfo {volume} {136}},\ \bibinfo {pages} {124107}
  (\bibinfo {year} {2012})},\ \Eprint
  {http://arxiv.org/abs/http://dx.doi.org/10.1063/1.3696967}
  {http://dx.doi.org/10.1063/1.3696967} \BibitemShut {NoStop}%
\bibitem [{\citenamefont {Ayers}\ and\ \citenamefont
  {Levy}(2009)}]{Ayers_2009}%
  \BibitemOpen
  \bibfield  {author} {\bibinfo {author} {\bibfnamefont {P.~W.}\ \bibnamefont
  {Ayers}}\ and\ \bibinfo {author} {\bibfnamefont {M.}~\bibnamefont {Levy}},\
  }\href {\doibase 10.1103/physreva.80.012508} {\bibfield  {journal} {\bibinfo
  {journal} {{Phys. Rev. A}}\ }\textbf {\bibinfo {volume} {80}} (\bibinfo
  {year} {2009}),\ 10.1103/physreva.80.012508}\BibitemShut {NoStop}%
\bibitem [{\citenamefont {Ayers}\ \emph {et~al.}(2012)\citenamefont {Ayers},
  \citenamefont {Levy},\ and\ \citenamefont {Nagy}}]{Ayers_2012}%
  \BibitemOpen
  \bibfield  {author} {\bibinfo {author} {\bibfnamefont {P.~W.}\ \bibnamefont
  {Ayers}}, \bibinfo {author} {\bibfnamefont {M.}~\bibnamefont {Levy}}, \ and\
  \bibinfo {author} {\bibfnamefont {A.}~\bibnamefont {Nagy}},\ }\href {\doibase
  10.1103/physreva.85.042518} {\bibfield  {journal} {\bibinfo  {journal}
  {{Phys. Rev. A}}\ }\textbf {\bibinfo {volume} {85}} (\bibinfo {year}
  {2012}),\ 10.1103/physreva.85.042518}\BibitemShut {NoStop}%
\bibitem [{\citenamefont {Levy}\ and\ \citenamefont {Nagy}(1999)}]{Levy_1999}%
  \BibitemOpen
  \bibfield  {author} {\bibinfo {author} {\bibfnamefont {M.}~\bibnamefont
  {Levy}}\ and\ \bibinfo {author} {\bibfnamefont {{\'{A}}.}~\bibnamefont
  {Nagy}},\ }\href {\doibase 10.1103/physrevlett.83.4361} {\bibfield  {journal}
  {\bibinfo  {journal} {{Phys. Rev. Lett.}}\ }\textbf {\bibinfo {volume}
  {83}},\ \bibinfo {pages} {4361} (\bibinfo {year} {1999})}\BibitemShut
  {NoStop}%
\bibitem [{\citenamefont {Baroni}\ \emph {et~al.}(1987)\citenamefont {Baroni},
  \citenamefont {Giannozzi},\ and\ \citenamefont {Testa}}]{Baroni_1987}%
  \BibitemOpen
  \bibfield  {author} {\bibinfo {author} {\bibfnamefont {S.}~\bibnamefont
  {Baroni}}, \bibinfo {author} {\bibfnamefont {P.}~\bibnamefont {Giannozzi}}, \
  and\ \bibinfo {author} {\bibfnamefont {A.}~\bibnamefont {Testa}},\ }\href
  {\doibase 10.1103/physrevlett.58.1861} {\bibfield  {journal} {\bibinfo
  {journal} {{Phys. Rev. Lett.}}\ }\textbf {\bibinfo {volume} {58}},\ \bibinfo
  {pages} {1861} (\bibinfo {year} {1987})}\BibitemShut {NoStop}%
\bibitem [{\citenamefont {Lieb}(1983)}]{Lieb_1983}%
  \BibitemOpen
  \bibfield  {author} {\bibinfo {author} {\bibfnamefont {E.~H.}\ \bibnamefont
  {Lieb}},\ }\href {\doibase 10.1002/qua.560240302} {\bibfield  {journal}
  {\bibinfo  {journal} {{Int. J. Quantum Chem.}}\ }\textbf {\bibinfo {volume}
  {24}},\ \bibinfo {pages} {243} (\bibinfo {year} {1983})}\BibitemShut
  {NoStop}%
\bibitem [{\citenamefont {Schipper}\ \emph {et~al.}(1998)\citenamefont
  {Schipper}, \citenamefont {Gritsenko},\ and\ \citenamefont
  {Baerends}}]{Schipper_1998}%
  \BibitemOpen
  \bibfield  {author} {\bibinfo {author} {\bibfnamefont {P.~R.~T.}\
  \bibnamefont {Schipper}}, \bibinfo {author} {\bibfnamefont {O.~V.}\
  \bibnamefont {Gritsenko}}, \ and\ \bibinfo {author} {\bibfnamefont {E.~J.}\
  \bibnamefont {Baerends}},\ }\href {\doibase 10.1007/s002140050343} {\bibfield
   {journal} {\bibinfo  {journal} {Theor. Chem. Acc.}\ }\textbf {\bibinfo
  {volume} {99}},\ \bibinfo {pages} {329} (\bibinfo {year} {1998})}\BibitemShut
  {NoStop}%
\bibitem [{\citenamefont {Gritsenko}\ and\ \citenamefont
  {Baerends}(1997)}]{Gritsenko_1997}%
  \BibitemOpen
  \bibfield  {author} {\bibinfo {author} {\bibfnamefont {O.~V.}\ \bibnamefont
  {Gritsenko}}\ and\ \bibinfo {author} {\bibfnamefont {E.~J.}\ \bibnamefont
  {Baerends}},\ }\href {\doibase 10.1007/s002140050202} {\bibfield  {journal}
  {\bibinfo  {journal} {Theor. Chem. Acc.}\ }\textbf {\bibinfo {volume} {96}},\
  \bibinfo {pages} {44} (\bibinfo {year} {1997})}\BibitemShut {NoStop}%
\bibitem [{\citenamefont {Wang}\ and\ \citenamefont
  {Schwarz}(1996)}]{Wang_1996}%
  \BibitemOpen
  \bibfield  {author} {\bibinfo {author} {\bibfnamefont {S.~G.}\ \bibnamefont
  {Wang}}\ and\ \bibinfo {author} {\bibfnamefont {W.~H.~E.}\ \bibnamefont
  {Schwarz}},\ }\href {\doibase 10.1063/1.472307} {\bibfield  {journal}
  {\bibinfo  {journal} {{J. Chem. Phys.}}\ }\textbf {\bibinfo {volume} {105}},\
  \bibinfo {pages} {4641} (\bibinfo {year} {1996})}\BibitemShut {NoStop}%
\bibitem [{\citenamefont {Slater}\ \emph {et~al.}(1969)\citenamefont {Slater},
  \citenamefont {Mann}, \citenamefont {Wilson},\ and\ \citenamefont
  {Wood}}]{Slater_1969}%
  \BibitemOpen
  \bibfield  {author} {\bibinfo {author} {\bibfnamefont {J.~C.}\ \bibnamefont
  {Slater}}, \bibinfo {author} {\bibfnamefont {J.~B.}\ \bibnamefont {Mann}},
  \bibinfo {author} {\bibfnamefont {T.~M.}\ \bibnamefont {Wilson}}, \ and\
  \bibinfo {author} {\bibfnamefont {J.~H.}\ \bibnamefont {Wood}},\ }\href
  {\doibase 10.1103/physrev.184.672} {\bibfield  {journal} {\bibinfo  {journal}
  {Physical Review}\ }\textbf {\bibinfo {volume} {184}},\ \bibinfo {pages}
  {672} (\bibinfo {year} {1969})}\BibitemShut {NoStop}%
\bibitem [{\citenamefont {Baerends}\ \emph {et~al.}(2019)\citenamefont
  {Baerends}, \citenamefont {Ziegler}, \citenamefont {Atkins}, \citenamefont
  {Autschbach}, \citenamefont {Bashford}, \citenamefont {Baseggio},
  \citenamefont {B{\'{e}}rces}, \citenamefont {Bickelhaupt}, \citenamefont
  {Bo}, \citenamefont {Boerritger}, \citenamefont {Cavallo}, \citenamefont
  {Daul}, \citenamefont {Chong}, \citenamefont {Chulhai}, \citenamefont {Deng},
  \citenamefont {Dickson}, \citenamefont {Dieterich}, \citenamefont {Ellis},
  \citenamefont {van Faassen}, \citenamefont {Ghysels}, \citenamefont
  {Giammona}, \citenamefont {van Gisbergen}, \citenamefont {Goez},
  \citenamefont {G{\"{o}}tz}, \citenamefont {Gusarov}, \citenamefont {Harris},
  \citenamefont {van~den Hoek}, \citenamefont {Hu}, \citenamefont {Jacob},
  \citenamefont {Jacobsen}, \citenamefont {Jensen}, \citenamefont {Joubert},
  \citenamefont {Kaminski}, \citenamefont {van Kessel}, \citenamefont
  {K{\"{o}}nig}, \citenamefont {Kootstra}, \citenamefont {Kovalenko},
  \citenamefont {Krykunov}, \citenamefont {van Lenthe}, \citenamefont
  {McCormack}, \citenamefont {Michalak}, \citenamefont {Mitoraj}, \citenamefont
  {Morton}, \citenamefont {Neugebauer}, \citenamefont {Nicu}, \citenamefont
  {Noodleman}, \citenamefont {Osinga}, \citenamefont {Patchkovskii},
  \citenamefont {Pavanello}, \citenamefont {Peeples}, \citenamefont
  {Philipsen}, \citenamefont {Post}, \citenamefont {Pye}, \citenamefont
  {Ramanantoanina}, \citenamefont {Ramos}, \citenamefont {Ravenek},
  \citenamefont {Rodr{\'{i}}guez}, \citenamefont {Ros}, \citenamefont
  {R{\"{u}}ger}, \citenamefont {Schipper}, \citenamefont {Schl{\"{u}}ns},
  \citenamefont {van Schoot}, \citenamefont {Schreckenbach}, \citenamefont
  {Seldenthuis}, \citenamefont {Seth}, \citenamefont {Snijders}, \citenamefont
  {Sol{\`{a}}}, \citenamefont {M.}, \citenamefont {Swart}, \citenamefont
  {Swerhone}, \citenamefont {te~Velde}, \citenamefont {Tognetti}, \citenamefont
  {Vernooijs}, \citenamefont {Versluis}, \citenamefont {Visscher},
  \citenamefont {Visser}, \citenamefont {Wang}, \citenamefont {Wesolowski},
  \citenamefont {van Wezenbeek}, \citenamefont {Wiesenekker}, \citenamefont
  {Wolff}, \citenamefont {Woo},\ and\ \citenamefont
  {Yakovlev}}]{ADF2019authors}%
  \BibitemOpen
  \bibfield  {author} {\bibinfo {author} {\bibfnamefont {E.~J.}\ \bibnamefont
  {Baerends}}, \bibinfo {author} {\bibfnamefont {T.}~\bibnamefont {Ziegler}},
  \bibinfo {author} {\bibfnamefont {A.~J.}\ \bibnamefont {Atkins}}, \bibinfo
  {author} {\bibfnamefont {J.}~\bibnamefont {Autschbach}}, \bibinfo {author}
  {\bibfnamefont {D.}~\bibnamefont {Bashford}}, \bibinfo {author}
  {\bibfnamefont {O.}~\bibnamefont {Baseggio}}, \bibinfo {author}
  {\bibfnamefont {A.}~\bibnamefont {B{\'{e}}rces}}, \bibinfo {author}
  {\bibfnamefont {F.~M.}\ \bibnamefont {Bickelhaupt}}, \bibinfo {author}
  {\bibfnamefont {C.}~\bibnamefont {Bo}}, \bibinfo {author} {\bibfnamefont
  {P.~M.}\ \bibnamefont {Boerritger}}, \bibinfo {author} {\bibfnamefont
  {L.}~\bibnamefont {Cavallo}}, \bibinfo {author} {\bibfnamefont
  {C.}~\bibnamefont {Daul}}, \bibinfo {author} {\bibfnamefont {D.~P.}\
  \bibnamefont {Chong}}, \bibinfo {author} {\bibfnamefont {D.~V.}\ \bibnamefont
  {Chulhai}}, \bibinfo {author} {\bibfnamefont {L.}~\bibnamefont {Deng}},
  \bibinfo {author} {\bibfnamefont {R.~M.}\ \bibnamefont {Dickson}}, \bibinfo
  {author} {\bibfnamefont {J.~M.}\ \bibnamefont {Dieterich}}, \bibinfo {author}
  {\bibfnamefont {D.~E.}\ \bibnamefont {Ellis}}, \bibinfo {author}
  {\bibfnamefont {M.}~\bibnamefont {van Faassen}}, \bibinfo {author}
  {\bibfnamefont {A.}~\bibnamefont {Ghysels}}, \bibinfo {author} {\bibfnamefont
  {A.}~\bibnamefont {Giammona}}, \bibinfo {author} {\bibfnamefont {S.~J.~A.}\
  \bibnamefont {van Gisbergen}}, \bibinfo {author} {\bibfnamefont
  {A.}~\bibnamefont {Goez}}, \bibinfo {author} {\bibfnamefont {A.~W.}\
  \bibnamefont {G{\"{o}}tz}}, \bibinfo {author} {\bibfnamefont
  {S.}~\bibnamefont {Gusarov}}, \bibinfo {author} {\bibfnamefont {F.~E.}\
  \bibnamefont {Harris}}, \bibinfo {author} {\bibfnamefont {P.}~\bibnamefont
  {van~den Hoek}}, \bibinfo {author} {\bibfnamefont {Z.}~\bibnamefont {Hu}},
  \bibinfo {author} {\bibfnamefont {C.~R.}\ \bibnamefont {Jacob}}, \bibinfo
  {author} {\bibfnamefont {H.}~\bibnamefont {Jacobsen}}, \bibinfo {author}
  {\bibfnamefont {L.}~\bibnamefont {Jensen}}, \bibinfo {author} {\bibfnamefont
  {L.}~\bibnamefont {Joubert}}, \bibinfo {author} {\bibfnamefont {J.~W.}\
  \bibnamefont {Kaminski}}, \bibinfo {author} {\bibfnamefont {G.}~\bibnamefont
  {van Kessel}}, \bibinfo {author} {\bibfnamefont {C.}~\bibnamefont
  {K{\"{o}}nig}}, \bibinfo {author} {\bibfnamefont {F.}~\bibnamefont
  {Kootstra}}, \bibinfo {author} {\bibfnamefont {A.}~\bibnamefont {Kovalenko}},
  \bibinfo {author} {\bibfnamefont {M.}~\bibnamefont {Krykunov}}, \bibinfo
  {author} {\bibfnamefont {E.}~\bibnamefont {van Lenthe}}, \bibinfo {author}
  {\bibfnamefont {D.~A.}\ \bibnamefont {McCormack}}, \bibinfo {author}
  {\bibfnamefont {A.}~\bibnamefont {Michalak}}, \bibinfo {author}
  {\bibfnamefont {M.}~\bibnamefont {Mitoraj}}, \bibinfo {author} {\bibfnamefont
  {S.~M.}\ \bibnamefont {Morton}}, \bibinfo {author} {\bibfnamefont
  {J.}~\bibnamefont {Neugebauer}}, \bibinfo {author} {\bibfnamefont {V.~P.}\
  \bibnamefont {Nicu}}, \bibinfo {author} {\bibfnamefont {L.}~\bibnamefont
  {Noodleman}}, \bibinfo {author} {\bibfnamefont {V.~P.}\ \bibnamefont
  {Osinga}}, \bibinfo {author} {\bibfnamefont {S.}~\bibnamefont
  {Patchkovskii}}, \bibinfo {author} {\bibfnamefont {M.}~\bibnamefont
  {Pavanello}}, \bibinfo {author} {\bibfnamefont {C.~A.}\ \bibnamefont
  {Peeples}}, \bibinfo {author} {\bibfnamefont {P.~H.~T.}\ \bibnamefont
  {Philipsen}}, \bibinfo {author} {\bibfnamefont {D.}~\bibnamefont {Post}},
  \bibinfo {author} {\bibfnamefont {C.~C.}\ \bibnamefont {Pye}}, \bibinfo
  {author} {\bibfnamefont {H.}~\bibnamefont {Ramanantoanina}}, \bibinfo
  {author} {\bibfnamefont {P.}~\bibnamefont {Ramos}}, \bibinfo {author}
  {\bibfnamefont {W.}~\bibnamefont {Ravenek}}, \bibinfo {author} {\bibfnamefont
  {J.~I.}\ \bibnamefont {Rodr{\'{i}}guez}}, \bibinfo {author} {\bibfnamefont
  {P.}~\bibnamefont {Ros}}, \bibinfo {author} {\bibfnamefont {R.}~\bibnamefont
  {R{\"{u}}ger}}, \bibinfo {author} {\bibfnamefont {P.~R.~T.}\ \bibnamefont
  {Schipper}}, \bibinfo {author} {\bibfnamefont {D.}~\bibnamefont
  {Schl{\"{u}}ns}}, \bibinfo {author} {\bibfnamefont {H.}~\bibnamefont {van
  Schoot}}, \bibinfo {author} {\bibfnamefont {G.}~\bibnamefont
  {Schreckenbach}}, \bibinfo {author} {\bibfnamefont {J.~S.}\ \bibnamefont
  {Seldenthuis}}, \bibinfo {author} {\bibfnamefont {M.}~\bibnamefont {Seth}},
  \bibinfo {author} {\bibfnamefont {J.~G.}\ \bibnamefont {Snijders}}, \bibinfo
  {author} {\bibfnamefont {M.}~\bibnamefont {Sol{\`{a}}}}, \bibinfo {author}
  {\bibfnamefont {S.}~\bibnamefont {M.}}, \bibinfo {author} {\bibfnamefont
  {M.}~\bibnamefont {Swart}}, \bibinfo {author} {\bibfnamefont
  {D.}~\bibnamefont {Swerhone}}, \bibinfo {author} {\bibfnamefont
  {G.}~\bibnamefont {te~Velde}}, \bibinfo {author} {\bibfnamefont
  {V.}~\bibnamefont {Tognetti}}, \bibinfo {author} {\bibfnamefont
  {P.}~\bibnamefont {Vernooijs}}, \bibinfo {author} {\bibfnamefont
  {L.}~\bibnamefont {Versluis}}, \bibinfo {author} {\bibfnamefont
  {L.}~\bibnamefont {Visscher}}, \bibinfo {author} {\bibfnamefont
  {O.}~\bibnamefont {Visser}}, \bibinfo {author} {\bibfnamefont
  {F.}~\bibnamefont {Wang}}, \bibinfo {author} {\bibfnamefont {T.~A.}\
  \bibnamefont {Wesolowski}}, \bibinfo {author} {\bibfnamefont {E.~M.}\
  \bibnamefont {van Wezenbeek}}, \bibinfo {author} {\bibfnamefont
  {G.}~\bibnamefont {Wiesenekker}}, \bibinfo {author} {\bibfnamefont {S.~K.}\
  \bibnamefont {Wolff}}, \bibinfo {author} {\bibfnamefont {T.~K.}\ \bibnamefont
  {Woo}}, \ and\ \bibinfo {author} {\bibfnamefont {A.~L.}\ \bibnamefont
  {Yakovlev}},\ }\href@noop {} {\enquote {\bibinfo {title} {{ADF2019, SCM,
  Theoretical Chemistry, Vrije Universiteit, Amsterdam, The Netherlands,
  https://www.scm.com}},}\ } (\bibinfo {year} {2019})\BibitemShut {NoStop}%
\bibitem [{\citenamefont {Perdew}\ \emph {et~al.}(1996)\citenamefont {Perdew},
  \citenamefont {Burke},\ and\ \citenamefont {Ernzerhof}}]{PBEc}%
  \BibitemOpen
  \bibfield  {author} {\bibinfo {author} {\bibfnamefont {J.~P.}\ \bibnamefont
  {Perdew}}, \bibinfo {author} {\bibfnamefont {K.}~\bibnamefont {Burke}}, \
  and\ \bibinfo {author} {\bibfnamefont {M.}~\bibnamefont {Ernzerhof}},\ }\href
  {\doibase 10.1103/PhysRevLett.77.3865} {\bibfield  {journal} {\bibinfo
  {journal} {{Phys. Rev. Lett.}}\ }\textbf {\bibinfo {volume} {77}},\ \bibinfo
  {pages} {3865} (\bibinfo {year} {1996})}\BibitemShut {NoStop}%
\bibitem [{\citenamefont {Zhao}\ and\ \citenamefont
  {Truhlar}(2008)}]{zhao2008}%
  \BibitemOpen
  \bibfield  {author} {\bibinfo {author} {\bibfnamefont {Y.}~\bibnamefont
  {Zhao}}\ and\ \bibinfo {author} {\bibfnamefont {D.~G.}\ \bibnamefont
  {Truhlar}},\ }\href@noop {} {\bibfield  {journal} {\bibinfo  {journal}
  {{Theor. Chem. Acc.}}\ }\textbf {\bibinfo {volume} {120}},\ \bibinfo {pages}
  {215} (\bibinfo {year} {2008})}\BibitemShut {NoStop}%
\bibitem [{\citenamefont {Sun}\ \emph {et~al.}(2015)\citenamefont {Sun},
  \citenamefont {Ruzsinszky},\ and\ \citenamefont {Perdew}}]{Sun_2015}%
  \BibitemOpen
  \bibfield  {author} {\bibinfo {author} {\bibfnamefont {J.}~\bibnamefont
  {Sun}}, \bibinfo {author} {\bibfnamefont {A.}~\bibnamefont {Ruzsinszky}}, \
  and\ \bibinfo {author} {\bibfnamefont {J.~P.}\ \bibnamefont {Perdew}},\
  }\href {\doibase 10.1103/physrevlett.115.036402} {\bibfield  {journal}
  {\bibinfo  {journal} {{Phys. Rev. Lett.}}\ }\textbf {\bibinfo {volume} {115}}
  (\bibinfo {year} {2015}),\ 10.1103/physrevlett.115.036402}\BibitemShut
  {NoStop}%
\bibitem [{\citenamefont {Perdew}\ \emph {et~al.}(2009)\citenamefont {Perdew},
  \citenamefont {Ruzsinszky}, \citenamefont {Csonka}, \citenamefont
  {Constantin},\ and\ \citenamefont {Sun}}]{PErdew_2009}%
  \BibitemOpen
  \bibfield  {author} {\bibinfo {author} {\bibfnamefont {J.~P.}\ \bibnamefont
  {Perdew}}, \bibinfo {author} {\bibfnamefont {A.}~\bibnamefont {Ruzsinszky}},
  \bibinfo {author} {\bibfnamefont {G.~I.}\ \bibnamefont {Csonka}}, \bibinfo
  {author} {\bibfnamefont {L.~A.}\ \bibnamefont {Constantin}}, \ and\ \bibinfo
  {author} {\bibfnamefont {J.}~\bibnamefont {Sun}},\ }\href {\doibase
  10.1103/physrevlett.103.026403} {\bibfield  {journal} {\bibinfo  {journal}
  {{Phys. Rev. Lett.}}\ }\textbf {\bibinfo {volume} {103}} (\bibinfo {year}
  {2009}),\ 10.1103/physrevlett.103.026403}\BibitemShut {NoStop}%
\bibitem [{\citenamefont {Neese}(2012)}]{neese2012}%
  \BibitemOpen
  \bibfield  {author} {\bibinfo {author} {\bibfnamefont {F.}~\bibnamefont
  {Neese}},\ }\href {\doibase 10.1002/wcms.81} {\bibfield  {journal} {\bibinfo
  {journal} {WIREs: Comput. Mol. Sci.}\ }\textbf {\bibinfo {volume} {2}},\
  \bibinfo {pages} {73} (\bibinfo {year} {2012})}\BibitemShut {NoStop}%
\bibitem [{\citenamefont {Frisch}\ \emph {et~al.}(2016)\citenamefont {Frisch},
  \citenamefont {Trucks}, \citenamefont {Schlegel}, \citenamefont {Scuseria},
  \citenamefont {Robb}, \citenamefont {Cheeseman}, \citenamefont {Scalmani},
  \citenamefont {Barone}, \citenamefont {Petersson}, \citenamefont {Nakatsuji},
  \citenamefont {Li}, \citenamefont {Caricato}, \citenamefont {Marenich},
  \citenamefont {Bloino}, \citenamefont {Janesko}, \citenamefont {Gomperts},
  \citenamefont {Mennucci}, \citenamefont {Hratchian}, \citenamefont {Ortiz},
  \citenamefont {Izmaylov}, \citenamefont {Sonnenberg}, \citenamefont
  {Williams-Young}, \citenamefont {Ding}, \citenamefont {Lipparini},
  \citenamefont {Egidi}, \citenamefont {Goings}, \citenamefont {Peng},
  \citenamefont {Petrone}, \citenamefont {Henderson}, \citenamefont
  {Ranasinghe}, \citenamefont {Zakrzewski}, \citenamefont {Gao}, \citenamefont
  {Rega}, \citenamefont {Zheng}, \citenamefont {Liang}, \citenamefont {Hada},
  \citenamefont {Ehara}, \citenamefont {Toyota}, \citenamefont {Fukuda},
  \citenamefont {Hasegawa}, \citenamefont {Ishida}, \citenamefont {Nakajima},
  \citenamefont {Honda}, \citenamefont {Kitao}, \citenamefont {Nakai},
  \citenamefont {Vreven}, \citenamefont {Throssell}, \citenamefont
  {Montgomery}, \citenamefont {Peralta}, \citenamefont {Ogliaro}, \citenamefont
  {Bearpark}, \citenamefont {Heyd}, \citenamefont {Brothers}, \citenamefont
  {Kudin}, \citenamefont {Staroverov}, \citenamefont {Keith}, \citenamefont
  {Kobayashi}, \citenamefont {Normand}, \citenamefont {Raghavachari},
  \citenamefont {Rendell}, \citenamefont {Burant}, \citenamefont {Iyengar},
  \citenamefont {Tomasi}, \citenamefont {Cossi}, \citenamefont {Millam},
  \citenamefont {Klene}, \citenamefont {Adamo}, \citenamefont {Cammi},
  \citenamefont {Ochterski}, \citenamefont {Martin}, \citenamefont {Morokuma},
  \citenamefont {Farkas}, \citenamefont {Foresman},\ and\ \citenamefont
  {Fox}}]{g16}%
  \BibitemOpen
  \bibfield  {author} {\bibinfo {author} {\bibfnamefont {M.~J.}\ \bibnamefont
  {Frisch}}, \bibinfo {author} {\bibfnamefont {G.~W.}\ \bibnamefont {Trucks}},
  \bibinfo {author} {\bibfnamefont {H.~B.}\ \bibnamefont {Schlegel}}, \bibinfo
  {author} {\bibfnamefont {G.~E.}\ \bibnamefont {Scuseria}}, \bibinfo {author}
  {\bibfnamefont {M.~A.}\ \bibnamefont {Robb}}, \bibinfo {author}
  {\bibfnamefont {J.~R.}\ \bibnamefont {Cheeseman}}, \bibinfo {author}
  {\bibfnamefont {G.}~\bibnamefont {Scalmani}}, \bibinfo {author}
  {\bibfnamefont {V.}~\bibnamefont {Barone}}, \bibinfo {author} {\bibfnamefont
  {G.~A.}\ \bibnamefont {Petersson}}, \bibinfo {author} {\bibfnamefont
  {H.}~\bibnamefont {Nakatsuji}}, \bibinfo {author} {\bibfnamefont
  {X.}~\bibnamefont {Li}}, \bibinfo {author} {\bibfnamefont {M.}~\bibnamefont
  {Caricato}}, \bibinfo {author} {\bibfnamefont {A.~V.}\ \bibnamefont
  {Marenich}}, \bibinfo {author} {\bibfnamefont {J.}~\bibnamefont {Bloino}},
  \bibinfo {author} {\bibfnamefont {B.~G.}\ \bibnamefont {Janesko}}, \bibinfo
  {author} {\bibfnamefont {R.}~\bibnamefont {Gomperts}}, \bibinfo {author}
  {\bibfnamefont {B.}~\bibnamefont {Mennucci}}, \bibinfo {author}
  {\bibfnamefont {H.~P.}\ \bibnamefont {Hratchian}}, \bibinfo {author}
  {\bibfnamefont {J.~V.}\ \bibnamefont {Ortiz}}, \bibinfo {author}
  {\bibfnamefont {A.~F.}\ \bibnamefont {Izmaylov}}, \bibinfo {author}
  {\bibfnamefont {J.~L.}\ \bibnamefont {Sonnenberg}}, \bibinfo {author}
  {\bibfnamefont {D.}~\bibnamefont {Williams-Young}}, \bibinfo {author}
  {\bibfnamefont {F.}~\bibnamefont {Ding}}, \bibinfo {author} {\bibfnamefont
  {F.}~\bibnamefont {Lipparini}}, \bibinfo {author} {\bibfnamefont
  {F.}~\bibnamefont {Egidi}}, \bibinfo {author} {\bibfnamefont
  {J.}~\bibnamefont {Goings}}, \bibinfo {author} {\bibfnamefont
  {B.}~\bibnamefont {Peng}}, \bibinfo {author} {\bibfnamefont {A.}~\bibnamefont
  {Petrone}}, \bibinfo {author} {\bibfnamefont {T.}~\bibnamefont {Henderson}},
  \bibinfo {author} {\bibfnamefont {D.}~\bibnamefont {Ranasinghe}}, \bibinfo
  {author} {\bibfnamefont {V.~G.}\ \bibnamefont {Zakrzewski}}, \bibinfo
  {author} {\bibfnamefont {J.}~\bibnamefont {Gao}}, \bibinfo {author}
  {\bibfnamefont {N.}~\bibnamefont {Rega}}, \bibinfo {author} {\bibfnamefont
  {G.}~\bibnamefont {Zheng}}, \bibinfo {author} {\bibfnamefont
  {W.}~\bibnamefont {Liang}}, \bibinfo {author} {\bibfnamefont
  {M.}~\bibnamefont {Hada}}, \bibinfo {author} {\bibfnamefont {M.}~\bibnamefont
  {Ehara}}, \bibinfo {author} {\bibfnamefont {K.}~\bibnamefont {Toyota}},
  \bibinfo {author} {\bibfnamefont {R.}~\bibnamefont {Fukuda}}, \bibinfo
  {author} {\bibfnamefont {J.}~\bibnamefont {Hasegawa}}, \bibinfo {author}
  {\bibfnamefont {M.}~\bibnamefont {Ishida}}, \bibinfo {author} {\bibfnamefont
  {T.}~\bibnamefont {Nakajima}}, \bibinfo {author} {\bibfnamefont
  {Y.}~\bibnamefont {Honda}}, \bibinfo {author} {\bibfnamefont
  {O.}~\bibnamefont {Kitao}}, \bibinfo {author} {\bibfnamefont
  {H.}~\bibnamefont {Nakai}}, \bibinfo {author} {\bibfnamefont
  {T.}~\bibnamefont {Vreven}}, \bibinfo {author} {\bibfnamefont
  {K.}~\bibnamefont {Throssell}}, \bibinfo {author} {\bibfnamefont {J.~A.}\
  \bibnamefont {Montgomery}, \bibfnamefont {{Jr.}}}, \bibinfo {author}
  {\bibfnamefont {J.~E.}\ \bibnamefont {Peralta}}, \bibinfo {author}
  {\bibfnamefont {F.}~\bibnamefont {Ogliaro}}, \bibinfo {author} {\bibfnamefont
  {M.~J.}\ \bibnamefont {Bearpark}}, \bibinfo {author} {\bibfnamefont {J.~J.}\
  \bibnamefont {Heyd}}, \bibinfo {author} {\bibfnamefont {E.~N.}\ \bibnamefont
  {Brothers}}, \bibinfo {author} {\bibfnamefont {K.~N.}\ \bibnamefont {Kudin}},
  \bibinfo {author} {\bibfnamefont {V.~N.}\ \bibnamefont {Staroverov}},
  \bibinfo {author} {\bibfnamefont {T.~A.}\ \bibnamefont {Keith}}, \bibinfo
  {author} {\bibfnamefont {R.}~\bibnamefont {Kobayashi}}, \bibinfo {author}
  {\bibfnamefont {J.}~\bibnamefont {Normand}}, \bibinfo {author} {\bibfnamefont
  {K.}~\bibnamefont {Raghavachari}}, \bibinfo {author} {\bibfnamefont {A.~P.}\
  \bibnamefont {Rendell}}, \bibinfo {author} {\bibfnamefont {J.~C.}\
  \bibnamefont {Burant}}, \bibinfo {author} {\bibfnamefont {S.~S.}\
  \bibnamefont {Iyengar}}, \bibinfo {author} {\bibfnamefont {J.}~\bibnamefont
  {Tomasi}}, \bibinfo {author} {\bibfnamefont {M.}~\bibnamefont {Cossi}},
  \bibinfo {author} {\bibfnamefont {J.~M.}\ \bibnamefont {Millam}}, \bibinfo
  {author} {\bibfnamefont {M.}~\bibnamefont {Klene}}, \bibinfo {author}
  {\bibfnamefont {C.}~\bibnamefont {Adamo}}, \bibinfo {author} {\bibfnamefont
  {R.}~\bibnamefont {Cammi}}, \bibinfo {author} {\bibfnamefont {J.~W.}\
  \bibnamefont {Ochterski}}, \bibinfo {author} {\bibfnamefont {R.~L.}\
  \bibnamefont {Martin}}, \bibinfo {author} {\bibfnamefont {K.}~\bibnamefont
  {Morokuma}}, \bibinfo {author} {\bibfnamefont {O.}~\bibnamefont {Farkas}},
  \bibinfo {author} {\bibfnamefont {J.~B.}\ \bibnamefont {Foresman}}, \ and\
  \bibinfo {author} {\bibfnamefont {D.~J.}\ \bibnamefont {Fox}},\ }\href@noop
  {} {\enquote {\bibinfo {title} {Gaussian 16 {R}evision {C}.01},}\ } (\bibinfo
  {year} {2016}),\ \bibinfo {note} {gaussian Inc. Wallingford CT}\BibitemShut
  {NoStop}%
\bibitem [{\citenamefont {Saha}\ \emph {et~al.}(2006)\citenamefont {Saha},
  \citenamefont {Ehara},\ and\ \citenamefont {Nakatsuji}}]{saha2006}%
  \BibitemOpen
  \bibfield  {author} {\bibinfo {author} {\bibfnamefont {B.}~\bibnamefont
  {Saha}}, \bibinfo {author} {\bibfnamefont {M.}~\bibnamefont {Ehara}}, \ and\
  \bibinfo {author} {\bibfnamefont {H.}~\bibnamefont {Nakatsuji}},\ }\href
  {\doibase 10.1063/1.2200344} {\bibfield  {journal} {\bibinfo  {journal} {{J.
  Chem. Phys.}}\ }\textbf {\bibinfo {volume} {125}},\ \bibinfo {pages} {014316}
  (\bibinfo {year} {2006})},\ \Eprint
  {http://arxiv.org/abs/https://doi.org/10.1063/1.2200344}
  {https://doi.org/10.1063/1.2200344} \BibitemShut {NoStop}%
\bibitem [{\citenamefont {Estevez-Fregoso}\ and\ \citenamefont
  {Hernandez-Trujillo}(2016)}]{est2016}%
  \BibitemOpen
  \bibfield  {author} {\bibinfo {author} {\bibfnamefont {M.}~\bibnamefont
  {Estevez-Fregoso}}\ and\ \bibinfo {author} {\bibfnamefont {J.}~\bibnamefont
  {Hernandez-Trujillo}},\ }\href {\doibase 10.1039/C5CP06993A} {\bibfield
  {journal} {\bibinfo  {journal} {Phys. Chem. Chem. Phys.}\ }\textbf {\bibinfo
  {volume} {18}},\ \bibinfo {pages} {11792} (\bibinfo {year}
  {2016})}\BibitemShut {NoStop}%
\bibitem [{\citenamefont {Schreiber}\ \emph {et~al.}(2008)\citenamefont
  {Schreiber}, \citenamefont {Silva-Junior}, \citenamefont {Sauer},\ and\
  \citenamefont {Thiel}}]{Schrei2008a}%
  \BibitemOpen
  \bibfield  {author} {\bibinfo {author} {\bibfnamefont {M.}~\bibnamefont
  {Schreiber}}, \bibinfo {author} {\bibfnamefont {M.~R.}\ \bibnamefont
  {Silva-Junior}}, \bibinfo {author} {\bibfnamefont {S.~P.~A.}\ \bibnamefont
  {Sauer}}, \ and\ \bibinfo {author} {\bibfnamefont {W.}~\bibnamefont
  {Thiel}},\ }\href {\doibase 10.1063/1.2889385} {\bibfield  {journal}
  {\bibinfo  {journal} {{J. Chem. Phys.}}\ }\textbf {\bibinfo {volume} {128}},\
  \bibinfo {pages} {134110} (\bibinfo {year} {2008})},\ \Eprint
  {http://arxiv.org/abs/http://dx.doi.org/10.1063/1.2889385}
  {http://dx.doi.org/10.1063/1.2889385} \BibitemShut {NoStop}%
\bibitem [{\citenamefont {Silva-Junior}\ \emph {et~al.}(2008)\citenamefont
  {Silva-Junior}, \citenamefont {Schreiber}, \citenamefont {Sauer},\ and\
  \citenamefont {Thiel}}]{Schrei2008b}%
  \BibitemOpen
  \bibfield  {author} {\bibinfo {author} {\bibfnamefont {M.~R.}\ \bibnamefont
  {Silva-Junior}}, \bibinfo {author} {\bibfnamefont {M.}~\bibnamefont
  {Schreiber}}, \bibinfo {author} {\bibfnamefont {S.~P.~A.}\ \bibnamefont
  {Sauer}}, \ and\ \bibinfo {author} {\bibfnamefont {W.}~\bibnamefont
  {Thiel}},\ }\href {\doibase 10.1063/1.2973541} {\bibfield  {journal}
  {\bibinfo  {journal} {{J. Chem. Phys.}}\ }\textbf {\bibinfo {volume} {129}},\
  \bibinfo {pages} {104103} (\bibinfo {year} {2008})},\ \Eprint
  {http://arxiv.org/abs/http://dx.doi.org/10.1063/1.2973541}
  {http://dx.doi.org/10.1063/1.2973541} \BibitemShut {NoStop}%
\bibitem [{\citenamefont {Martins}\ \emph {et~al.}(2009)\citenamefont
  {Martins}, \citenamefont {Ferreira-Rodrigues}, \citenamefont {Rodrigues},
  \citenamefont {de~Souza}, \citenamefont {Mason}, \citenamefont {Eden},
  \citenamefont {Duflot}, \citenamefont {Flament}, \citenamefont {Hoffmann},
  \citenamefont {Delwiche}, \citenamefont {Hubin-Franskin},\ and\ \citenamefont
  {Limao-Vieira}}]{Martins2009}%
  \BibitemOpen
  \bibfield  {author} {\bibinfo {author} {\bibfnamefont {G.}~\bibnamefont
  {Martins}}, \bibinfo {author} {\bibfnamefont {A.~M.}\ \bibnamefont
  {Ferreira-Rodrigues}}, \bibinfo {author} {\bibfnamefont {F.~N.}\ \bibnamefont
  {Rodrigues}}, \bibinfo {author} {\bibfnamefont {G.~G.~B.}\ \bibnamefont
  {de~Souza}}, \bibinfo {author} {\bibfnamefont {N.~J.}\ \bibnamefont {Mason}},
  \bibinfo {author} {\bibfnamefont {S.}~\bibnamefont {Eden}}, \bibinfo {author}
  {\bibfnamefont {D.}~\bibnamefont {Duflot}}, \bibinfo {author} {\bibfnamefont
  {J.-P.}\ \bibnamefont {Flament}}, \bibinfo {author} {\bibfnamefont {S.~V.}\
  \bibnamefont {Hoffmann}}, \bibinfo {author} {\bibfnamefont {J.}~\bibnamefont
  {Delwiche}}, \bibinfo {author} {\bibfnamefont {M.-J.}\ \bibnamefont
  {Hubin-Franskin}}, \ and\ \bibinfo {author} {\bibfnamefont {P.}~\bibnamefont
  {Limao-Vieira}},\ }\href {\doibase 10.1039/B916620C} {\bibfield  {journal}
  {\bibinfo  {journal} {Phys. Chem. Chem. Phys.}\ }\textbf {\bibinfo {volume}
  {11}},\ \bibinfo {pages} {11219} (\bibinfo {year} {2009})}\BibitemShut
  {NoStop}%
\bibitem [{\citenamefont {Holland}\ \emph {et~al.}(2014)\citenamefont
  {Holland}, \citenamefont {Trofimov}, \citenamefont {Seddon}, \citenamefont
  {Gromov}, \citenamefont {Korona}, \citenamefont {de~Oliveira}, \citenamefont
  {Archer}, \citenamefont {Joyeux},\ and\ \citenamefont {Nahon}}]{Holland2014}%
  \BibitemOpen
  \bibfield  {author} {\bibinfo {author} {\bibfnamefont {D.~M.~P.}\
  \bibnamefont {Holland}}, \bibinfo {author} {\bibfnamefont {A.~B.}\
  \bibnamefont {Trofimov}}, \bibinfo {author} {\bibfnamefont {E.~A.}\
  \bibnamefont {Seddon}}, \bibinfo {author} {\bibfnamefont {E.~V.}\
  \bibnamefont {Gromov}}, \bibinfo {author} {\bibfnamefont {T.}~\bibnamefont
  {Korona}}, \bibinfo {author} {\bibfnamefont {N.}~\bibnamefont {de~Oliveira}},
  \bibinfo {author} {\bibfnamefont {L.~E.}\ \bibnamefont {Archer}}, \bibinfo
  {author} {\bibfnamefont {D.}~\bibnamefont {Joyeux}}, \ and\ \bibinfo {author}
  {\bibfnamefont {L.}~\bibnamefont {Nahon}},\ }\href {\doibase
  10.1039/C4CP02420F} {\bibfield  {journal} {\bibinfo  {journal} {Phys. Chem.
  Chem. Phys.}\ }\textbf {\bibinfo {volume} {16}},\ \bibinfo {pages} {21629}
  (\bibinfo {year} {2014})}\BibitemShut {NoStop}%
\bibitem [{\citenamefont {Floris}\ \emph {et~al.}(2014)\citenamefont {Floris},
  \citenamefont {Filippi},\ and\ \citenamefont {Amovilli}}]{Floris2014}%
  \BibitemOpen
  \bibfield  {author} {\bibinfo {author} {\bibfnamefont {F.~M.}\ \bibnamefont
  {Floris}}, \bibinfo {author} {\bibfnamefont {C.}~\bibnamefont {Filippi}}, \
  and\ \bibinfo {author} {\bibfnamefont {C.}~\bibnamefont {Amovilli}},\ }\href
  {\doibase 10.1063/1.4861429} {\bibfield  {journal} {\bibinfo  {journal} {{J.
  Chem. Phys.}}\ }\textbf {\bibinfo {volume} {140}},\ \bibinfo {pages} {034109}
  (\bibinfo {year} {2014})},\ \Eprint
  {http://arxiv.org/abs/http://dx.doi.org/10.1063/1.4861429}
  {http://dx.doi.org/10.1063/1.4861429} \BibitemShut {NoStop}%
\bibitem [{\citenamefont {Zimmerman}\ \emph {et~al.}(2010)\citenamefont
  {Zimmerman}, \citenamefont {Zhang},\ and\ \citenamefont
  {Musgrave}}]{Zimmer2010}%
  \BibitemOpen
  \bibfield  {author} {\bibinfo {author} {\bibfnamefont {P.~M.}\ \bibnamefont
  {Zimmerman}}, \bibinfo {author} {\bibfnamefont {Z.}~\bibnamefont {Zhang}}, \
  and\ \bibinfo {author} {\bibfnamefont {C.~B.}\ \bibnamefont {Musgrave}},\
  }\href {\doibase http://dx.doi.org/10.1038/nchem.694} {\bibfield  {journal}
  {\bibinfo  {journal} {Nat. Chem.}\ }\textbf {\bibinfo {volume} {2}},\
  \bibinfo {pages} {648} (\bibinfo {year} {2010})}\BibitemShut {NoStop}%
\end{thebibliography}%
\end{document}